\documentclass[journal]{IEEEtran}
\usepackage{pifont}  
\usepackage{amsthm}  
\usepackage{amsmath,amsfonts}
\usepackage{orcidlink}
\usepackage{float} 
\usepackage[linesnumbered, ruled, lined,boxed,commentsnumbered, algo2e]{algorithm2e}
\usepackage{algorithm}
\usepackage{algpseudocode}
\usepackage{array}
\usepackage[caption=false,font=normalsize,labelfont=sf,textfont=sf]{subfig}
\usepackage{textcomp}
\usepackage{stfloats}
\usepackage{url}
\usepackage{verbatim}
\usepackage{graphicx}
\usepackage{cite}
\usepackage{xcolor}
\usepackage{pdflscape}
\usepackage{rotating} 
\usepackage{booktabs} 
\usepackage{changepage} 
\usepackage{amsmath}  
\usepackage{mathrsfs} 
\usepackage{amssymb}  
\usepackage{multirow} 
\usepackage{subcaption}  
\usepackage{threeparttable} 

\newtheorem{definition}{Definition}  

\begin{document}
\font\cmsyten=cmsy10 at 9pt

\title{Enhancing Deep Learning-Based Rotational-XOR Attacks on Lightweight Block Ciphers Simon32/64 and Simeck32/64}

\author{Chengcai Liu, Siwei Chen~\orcidlink{0000-0003-3428-7647}, Zejun Xiang, Shasha Zhang, and Xiangyong Zeng}

\markboth{Submit to IEEE Transactions on Information Forensics and Security}%
{Chengcai Liu \MakeLowercase{\textit{et al.}}: Enhancing Deep Learning-Based Rotational-XOR Attacks on Lightweight Block Ciphers Simon32/64 and Simeck32/64}


\maketitle

\begin{abstract}
At CRYPTO 2019, Gohr pioneered neural cryptanalysis by introducing differential-based neural distinguishers to attack Speck32/64, establishing a novel paradigm combining deep learning with differential cryptanalysis.
Since then, constructing neural distinguishers has become a significant approach to achieving the deep learning-based cryptanalysis for block ciphers. 
This paper advances rotational-XOR (RX) attacks through neural networks, focusing on optimizing distinguishers and presenting key-recovery attacks for the lightweight block ciphers Simon32/64 and Simeck32/64.
In particular, we first construct the fundamental data formats specially designed for training RX-neural distinguishers by refining the existing data formats for differential-neural distinguishers. Based on these data formats, we systematically identify optimal RX-differences with Hamming weights 1 and 2 that develop high-accuracy RX-neural distinguishers. Then, through innovative application of the bit sensitivity test, we achieve significant compression of data format without sacrificing the distinguisher accuracy. This optimization enables us to add more multi-ciphertext pairs into the data formats, further strengthening the performance of RX-neural distinguishers. As an application, we obtain 14- and 17-round RX-neural distinguishers for Simon32/64 and Simeck32/64, which improves the previous ones by 3 and 2 rounds, respectively.
In addition, we propose two novel techniques, 
\textit{key bit sensitivity test} and the \textit{joint wrong key response}, to tackle the challenge of applying Bayesian's key-recovery strategy to the target cipher that adopts nonlinear key schedule in the related-key setting without considering of weak-key space. By this, we can straightforwardly mount a 17-round key-recovery attack on Simeck32/64 based on the improved 16-round RX-nerual distinguisher. To the best of our knowledge, the presented RX-neural distinguishers outperform the state-of-the-art neural-based distinguishers for both Simon32/64 and Simeck32/64, and this is the first successful neural-based key-recovery attack for Simeck32/64 under the related-key setting. 
\end{abstract}


\section{Introduction}
\IEEEPARstart{W}{ith} the development of information technology, deep learning has become a key technology driving the intelligence of various industries. For example, deep learning is used for diagnosing diseases and analyzing medical images in the medical field and perceiving driving environments and planning driving paths in the automotive industry. Moreover, deep learning also plays an important role in strengthening information security. For example, deep learning is an effective technique frequently applied to the side-channel attacks~\cite{timon2018non,kim2019make,DBLP:journals/tifs/WuWKLPBP23,DBLP:journals/tifs/KimHKSH24} in cryptography. In theory, cryptography and machine learning are naturally linked fields~\cite{rivest1991cryptography} because many cryptography tasks can be naturally framed as learning tasks. Therefore, combining deep learning with classical cryptanalysis technology has great potential. However, deep learning based cryptanalysis, like differential cryptanalysis~\cite{biham1991differential}, linear cryptanalysis~\cite{matsui1993linear}, rotational-XOR (RX) cryptanalysis~\cite{ashur2016rotational} and so on, is not much.

At CRYPTO 2019, Gohr~\cite{gohr2019improving} combined deep learning and differential cryptanalysis to attack Speck32/64. He trained a model with training datasets of ciphertext pairs and labels. The model, called the differential-neural distinguisher, can return a confidence level between 0 and 1 for a given ciphertext pair, indicating whether the ciphertext pair is a real ciphertext pair or a random ciphertext pair. Moreover, to mount key-recovery attacks using neural distinguishers, this work applies a variant of Bayesian optimization to recovering information of key bits, known as the Bayesian key-recovery strategy. As a result, Gohr carried out the 11-round attacks on Speck32/64 based on an 8-round differential-neural distinguisher. Since then, studies on combining classical cryptanalysis methods with deep learning techniques have emerged, like linear-neural cryptanalysis~\cite{hou2020linear}, integral-neural cryptanalysis~\cite{zahednejad2022improved}, RX-neural cryptanalysis~\cite{palmierideep}, etc.
The common idea of these works is to construct the corresponding neural distinguishers according to the concrete attacking types. Similar to classical cryptanalysis, good distinguishers can lead to good attacks. That is to say, the higher the accuracy of the neural distinguishers are, the longer the attacking rounds or the higher the attacking success rates will be. Therefore, how to optimize the accuracy of neural distinguishers is extremely necessary in the deep learning based attacks.

\subsection{Related Works}
In 2023, Chen et al.~\cite{chen2023new} used multi-ciphertext to improve the accuracy of differential-neural distinguishers. In the same year, Gohr et al.~\cite{gohr2022assessment} proposed a fair comparison scheme between multi-ciphertext and other neural distinguishers. They indicated that it is feasible to improve neural distinguishers using multi-ciphertext because of the dependence between ciphertext pairs. Taking 10-round Simon32/64 as an example, the accuracy of the differential-neural distinguisher trained with 64 ciphertext pairs is 0.36 higher than that of the case without multi-ciphertext. 
 
At SAC 2023, Ebrahimi et al.~\cite{palmierideep} first introduced the RX-neural distinguisher. They utilized an evolutionary algorithm framework and obtained 11- and 15-round RX-neural distinguishers for Simon32/64 and Simeck32/64, respectively. This study marks the first attempt to construct an RX-neural distinguisher, but key-recovery attacks based on neural distinguishers were not discussed. We guess, the reason is that RX cryptanalysis is a related-key attack method, while the Bayesian key-recovery strategy is only applicable to single-key attacks. This is because the crucial step in the Bayesian key-recovery strategy involves performing \textit{the wrong key response} (WKR), which is used to update the single guessed key. However, how to perform WKR in a related-key setting remains unsolved.

At ASIACRYPT 2023, Bao et al.~\cite{bao2023more} successfully applied related-key attacks using the Bayesian key-recovery strategy within the weak-key space to Speck32/64. This work provides new insights into the application of the Bayesian key-recovery strategy in a related-key setting. However, the weak-key space, which requires the key to satisfy certain special conditions, significantly reduces the key space. In other words, the requirements for this attack scenario are quite strict.

\subsection{Motivation of This Work}
Indeed, the application of multi-ciphertext can significantly improve the accuracy of neural distinguishers, but the platform is required to have extremely powerful computational resources. In order to enhance the accuracy under the limited computing resources, the data format needs to be cut down as short as possible. In addition, to our best knowledge, there is no related work about deep learning based RX attacks besides Ebrahimi et al.'s work (SAC 2023). Moreover, Ebrahimi et al. presented the RX-neural distinguishers in a rather simple way, without optimizing them or mounting any key-recovery attacks. Furthermore, with regard to related-key attacks on ciphers with nonlinear key schedules, Bao et al.'s method effectively addresses the challenges posed by key propagation probabilities in the Bayesian key-recovery strategy. However, it also reduces the key space, which in turn impacts the success rate of the final key-recovery attack. Therefore, expanding the key space or entirely avoiding the introduction of weak keys is worth  further investigation in the context of related-key settings.

\subsection{Our Contributions}
Motivated by the existing problems as mentioned above, we in this paper focus on improving deep learning based RX attacks on Simon32/64 and Simeck32/64, and give the following contributions:

\vspace{5pt}
\noindent\textbf{Exploring the better RX-differences and fundamental data formats for training RX-neural distinguishers.} To train an RX-neural distinguisher, the first step is to collect ciphertext pairs using a fixed RX-difference and construct training dataset based on a given data format. Thus, it is quite crucial for training high-accuracy distinguisher to find good RX-differences as well as appropriate data formats. Investigated from the previous works about the classical as well as deep learning based RX-cryptanalysis, the RX-difference with lower Hamming weight is more likely to derive the better results. Besides, we adjust the existing data formats, which are proved to be effective for training differential-neural distinguishers, to fit RX-neural distinguishers training. Through exhaustively evaluating the RX-nerual distinguishers trained using all possible RX-differences with Hamming weights of 1 and 2, we respectively retain six good RX-differences for Simon32/64 and Simeck32/64. Among these, the highest accuracy for 11-round Simon32/64 and 13-round Simeck32/64 are 0.9215 and 0.6990 whereas the presented ones in~\cite{palmierideep} are 0.5445 and 0.7057, respectively. Notably, our results are obtained without employing any optimization training strategy but the authors in~\cite{palmierideep} leveraged the multi-cipher technique, demonstrating that our strategy for good RX-differences selection is indeed effective.

\vspace{5pt}
\noindent\textbf{Presenting currently best neural distinguishers by leveraging the multi-ciphertext and staged strategies.}
A critical strategy for enhancing neural distinguisher accuracy involves expanding data formats with multi-ciphertext. While incorporating additional ciphertext pairs improves model discriminative power, this approach proportionally escalates computational demands. 
To make training feasible within limited computational resources, we need to first cut down the fundamental data format as short as possible by identifying and eliminating the redundant components, then add into ciphertext pairs. For this purpose, we utilize the \textit{bit sensitivity test} (BST) technique, introduced by Chen et al.~\cite{chen2020neural}, to analyze the influence of each ciphertext bit on RX-neural distinguishers. Surprisingly, we find that the left branch of the ciphertext has negligible effect on the performance of distinguishers. In other words, the components related to the left branch of ciphertext are redundant. Thus, the number of components in the fundamental data can be optimized from 8 to 5. Moreover, for the remaining components, we use the method of controlling variables to gradually identify and eliminate those with minimal impact on accuracy. Consequently, we obtain two shortened data formats only with 2 and 3 components, respectively. Combining with the multi-ciphertext technique, we develop two multi-ciphertext data formats, successfully yielding 13- and 16-round distinguishers for Simon32/64 and Simeck32/64, respectively. Furthermore, by leveraging the staged training strategy, both the 13- and 16-round distinguishers are further improved by one round, which extend the state-of-art neural-based distinguishers from 13 to 14 for Simon32/64 and from 15 to 17 rounds for Simeck32/64. Our results, along with the various types of neural distinguishers are listed in Table~\ref{best_result}. 

\vspace{5pt}
\noindent\textbf{Mounting the first neural-based key-recovery attacks under related-key setting.} For Simon32/64, we can directly use Bayesian key-recovery strategy (BKS) as Simon adopts a linear key schedule. As a result, we achieve 14- and 15-round key-recovery attacks based on the presented RX-neural distinguishers, with success rates $100\%$ and $75\%$ respectively. 
To mount key-recovery attacks for Simeck32/64 without considering weak-key space, we introduce two novel techniques: the \textit{key bit sensitivity test} (KBST) and the \textit{joint wrong key response} (JWKR). Using KBST, we identify the key bits that have a significant impact on neural distinguishers, which are called sensitive key bits. Then, we only need to construct JWKR for the sensitive key bits instead of all key bits, which makes the key-guessing process practical. Combining with BKS, the key-recovery attacks can be achieved under the full key space in related-key setting. Based on the explored RX-nerual distinguishers, we conduct 16- and 17-round key-recovery attacks for Simeck32/64, with success rates $98\%$ and $40\%$. Compared to the existing neural-based key-recovery attacks (see Table~\ref{table:keyrecover_comparsion}), our results indeed seem weak as they make full use of the generic techniques of conventional attacks, which are not applicable to our attacks. Therefore, we can only consider to extend one round backward the distinguisher and directly utilize BKS. However, to our best knowledge, this is the first time to present key-recovery attacks based on neural distinguishers under related-key setting for Simon and Simeck. Also, the proposed KBST and JWKR are potential to neural-based related-key attacks for other block cipher like Speck.   
\begin{table}[!h]
		\caption{Comparison of our work with other studies on neural distinguishers for Simon32/64 and Simeck32/64}
              \renewcommand\tabcolsep{4pt}
			\begin{threeparttable}
				\begin{tabular}{ccccccc}
                     \toprule
					Cipher & Round & Accuracy  & TNR & TPR & Type\textsuperscript{$\dagger$} & Ref. \\ \midrule
                     \multirow{8}{*}{Simon32/64} & 11 & 0.5445 & - & - & RX & \cite{palmierideep} \\ 
                    & \textbf{11} & \textbf{1.0000} & \textbf{1.0000} & \textbf{1.0000} & RX & Sect. \ref{sec:4_2} \\ 

					& 12 & 0.6477 & 0.6518 & 0.6435 & RKD & \cite{lu2022improved} \\ 
                    & 12 & 0.5225 & - & - & SKD & \cite{zhang2022improving}  \\

     			    & 13 & 0.5262  & 0.5437 & 0.5081 & RKD & \cite{lu2022improved} \\ 
     			    & 13 & 0.5810 & 0.5730 & 0.5890 & RKD & \cite{yuan2025multi} \\ 
					& \textbf{13} & \textbf{0.7120} & \textbf{0.7075} & \textbf{0.7165} & RX & Sect. \ref{sec:4_2} \\ 
					& \textbf{14} & \textbf{0.5241} & \textbf{0.5634} & \textbf{0.4848} & RX & Sect. \ref{sec:4_2} \\ \midrule

					\multirow{7}{*}{Simeck32/64} & 12 & 0.5161 & 0.4807 & 0.5504 & SKD  & \cite{zhang2023improved} \\ 
                     & \textbf{12} & \textbf{1.0000}  & 1.0000 & 1.0000 & RX & Sect.~\ref{sec:4_2} \\ 
                    & 15 & 0.5467 & 0.5173 & 0.5762 & RKD & \cite{lu2022improved} \\ 
					& 15 & 0.5475 & - & - & RX & \cite{palmierideep} \\ 
                    & 15 & 0.5930 & 0.5950 & 0.5900 & RX & \cite{yuan2025multi} \\ 

					& \textbf{15} & \textbf{0.6042} & \textbf{0.5330} & \textbf{0.6750} & RX & Sect.~\ref{sec:4_2} \\   
					& \textbf{16} & \textbf{0.5130} & \textbf{0.5255} & \textbf{0.5000} & RX & Sect.~\ref{sec:4_2} \\ 
					& \textbf{17} & \textbf{0.5040} & \textbf{0.5950} & \textbf{0.4130} & RX & Sect.~\ref{sec:4_2} \\\bottomrule
				\end{tabular}
				
				\begin{tablenotes}
					\footnotesize
					\item[$\dagger$] RX: rotational-XOR; RKD: related-key differential; SKD: single-key differential;
				\end{tablenotes}
				
			\end{threeparttable}
		\label{best_result}
	\end{table}
\begin{table}[!h]
    \begin{center}
        \caption{Comparison of our and previous deep learning-based key-recovery attacks\label{table:keyrecover_comparsion}}
        \renewcommand\tabcolsep{2.7pt}
        \begin{threeparttable}

        \begin{tabular}{lllllccc}
            \toprule
            Cipher & Round$\dagger$ & Time & Data & Success rate & Type & Ref. &\\ \midrule
             \multirow{4}{*}{Simon32/64}  & 14{(13+1)} & $2^{32}$ & $2^{15.81}$ & 100\% & RX & Sect.~\ref{sec:5_0} \\
               & 15{(14+1)} & $2^{32}$ & $2^{15.81}$ & 75\% & RX & Sect.~\ref{sec:5_0} \\
              & 16{(4+11+1)} & $2^{42.79}$ & $2^{22}$ & 80\% & SKD &  \cite{zhang2022improving} \\
             
            & 17{(5+11+1)} & $2^{54.01}$ & $2^{28}$ & 9\% & SKD & \cite{zhang2022improving}  \\
                 
            \cmidrule{1-8}
            
             \multirow{5}{*}{Simeck32/64} & 15(3+10+2) & $2^{33.9}$ & $2^{24}$ & $88\%$ & SKD & \cite{DBLP:conf/isw/LyuTZ22} \\
             
             & 16{(4+11+1)}   &  $2^{24}$ & $2^{38.19}$ & 100\% & SKD &  \cite{zhang2023improved}\\
             
             & 16{(15+1)} & $2^{51}$ & $2^{16.17}$ & 98\% & RX &  Sect.~\ref{sec:6_0} \\
             & 17{(4+12+1)} & $2^{26}$ & $2^{45.04}$ & 30\% & SKD &  \cite{zhang2023improved}\\
             & 17{(16+1)} & $2^{54}$ & $2^{16.17}$ & 40\% & RX &  Sect.~\ref{sec:6_0} \\

            \bottomrule
        \end{tabular}
        
        \begin{tablenotes}
			\footnotesize
			\item[$\dagger$] The total attack rounds comprise the neural distinguisher rounds plus any prepended or appended rounds. For example, the 17-round attack on Simon32/64 prepends 5 rounds and appends 1 round to the 11-round neural distinguisher, denoted as 5+11+1.
        \end{tablenotes}
        \end{threeparttable}
    \end{center}
\end{table}

\subsection{Organization}
Sect.~\ref{sec:2_0} starts by giving some concepts and notations throughout this paper, then introduces Simon32/64, Simeck32/64 and conventional RX cryptanalysis in brief. We demonstrate how to identify good RX-differences in Sect.~\ref{sec:3_0} and optimize the prepared RX-nerual distinguishers from Sect.~\ref{sec:4_1} in Sect.~\ref{sec:4_0}. In Sect.~\ref{sec:5_0} and Sect.~\ref{sec:6_0}, we present the key-recovery attacks for Simon32/64 and Simeck32/64, repsectively. Finally, we conclude this paper in Sect.~\ref{sec:7_0}.

\section{Preliminary}
\label{sec:2_0}
\subsection{Notations and Concepts}
Simon and Simeck both adopt the Feistel structure, so we use $C_L$ and $C_R$ throughout this paper to denote the left and right branches of the ciphertext for ease. Also, some primary notations are given in Table~\ref{notation}.  
	\begin{table}[!h]
			\caption{Notations of this paper}
           \renewcommand\tabcolsep{13pt}
			\begin{tabular}{ll}
				\toprule
				Notations & Descriptions \\ 
                \midrule
				$\mathbb{F}_{2}$ & A finite field only contains 2 elements, i.e. 0 and 1\\
				$\mathbb{F}^{n}_{2}$ &  An $n$-dimensional vectorial space defined over $\mathbb{F}_{2}$\\
				$\odot$ & Bitwise AND \\
				$\oplus$ & Bitwise XOR\\
                $||$ & Concatenation of two bit-strings\\
				$x \lll \lambda$ & Circular left shift of $x$ by $\lambda$ bits\\ 
				$x \ggg \lambda$ & Circular right shift of $x$ by $\lambda$ bits\\
                $C^r_L, C^r_R$ & Left and right branches of $r$-round ciphertext\\
                $\Delta^r_L, \Delta^r_R$ & Left and right branches of $r$-round RX-difference\\
				\bottomrule
			\end{tabular}
			\label{notation}
	\end{table}

In addition, there are two key concepts, \textit{data format} and \textit{multi-ciphertext}, that will be frequently used in this paper. These concepts were mentioned in previous works but have not been formally defined. To ensure a clearer understanding, we provide their formal definitions. Besides, note that Simon and Simeck both adopt the Feistel structure, we use $C_L$ and $C_R$ throughout this paper to denote the left and right branches of the ciphertext for ease.

\begin{definition}[Data Format]
The \textit{data format} is a data tuple formed by $N (N\geq 1)$ elements $D_0,D_1,...,D_{N-1}$ used to generate the neural network training dataset, and denoted by
\[
    \mathscr{D} = (D_0,D_1,...,D_{N-1}),
\]
where each $D_i\in\mathbb{F}_2^{n}$ for $0\leq i \leq N-1$ corresponding to an $n$-bit data is called a \textit{component}. 
\end{definition}

For example, $\mathscr{D} = (C_L,C_R,C_L \oplus C'_L)$ is a data format that consists of three components where the corresponding plaintexts $(P,P')$ of $(C_L,C'_L)$ satisfies a deterministic relation. Specifically, the first and second components are the left and right branches of a ciphertext with size $2n$, respectively. The third component is the difference on the left branch of a pair of ciphertexts. Therefore, if a dataset is generated using $\mathscr{D}$, then each element in dataset has the same form as $(C_L,C_R,C_L \oplus C'_L)$ and is generated by one pair of ciphertexts. In order to improve the accuracy of neural distinguishers, in general, more pairs of ciphertexts can be appended into data format.

\begin{definition}[Multi-ciphertext Data Format]
Given a data format $\mathscr{D}=(D_0,...,D_{N-1})$. If there are $2k$ $(2k\leq N)$ different components $D_{k_i}(0\leq i\leq 2k-1)$ among $\{D_0,...,D_{N-1}\}$, and $(D_{k_{2i}},D_{k_{2i+1}})$ for all $0\leq i\leq k-1$ corresponds to a pair of ciphertexts, then we call $\mathscr{D}$ a $k$-\textit{multi-ciphertext data format} and specially denote it by $\mathscr{D}^{k}$.
\end{definition}

For example, $\mathscr{D}^3=(C_0, C_1, C_2, C_3, C'_0, C'_1, C'_2)$ is a 3-multi-ciphertext data format. In particular, the pair of components $(C_i,C'_{i})$ for $i\in\{0,1,2\}$ represents a pair of ciphertexts, where the corresponding pair of plaintexts $(P_i,P'_i)$ satisfies a deterministic relation. 
	
\subsection{Description of Simon and Simeck}
\begin{figure}[!th]
        \centering
        \subfloat[Simon32/64]{\includegraphics[width=1.6in]{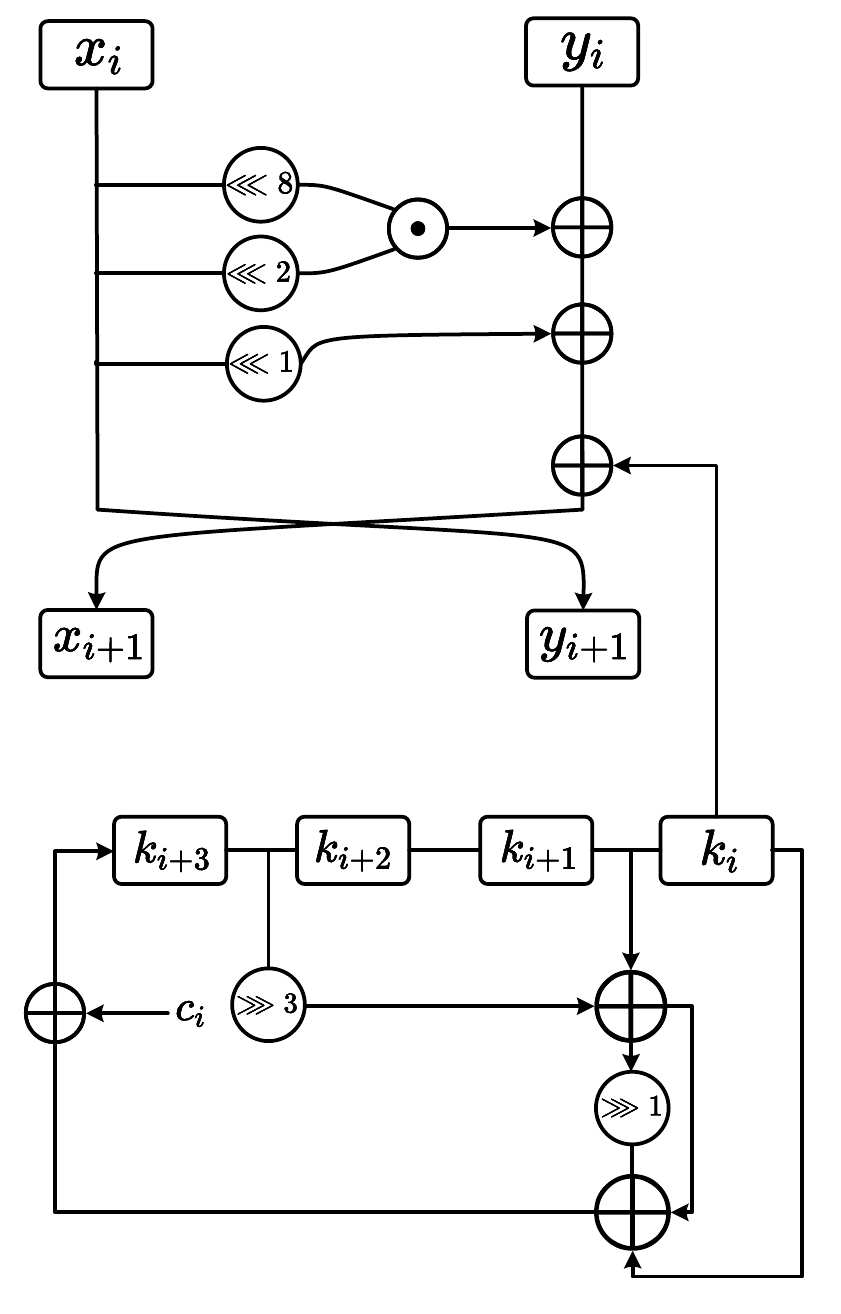}\label{fig:2simon_cipher}}
        \hfil
        \subfloat[Simeck32/64]{\includegraphics[width=1.7in]{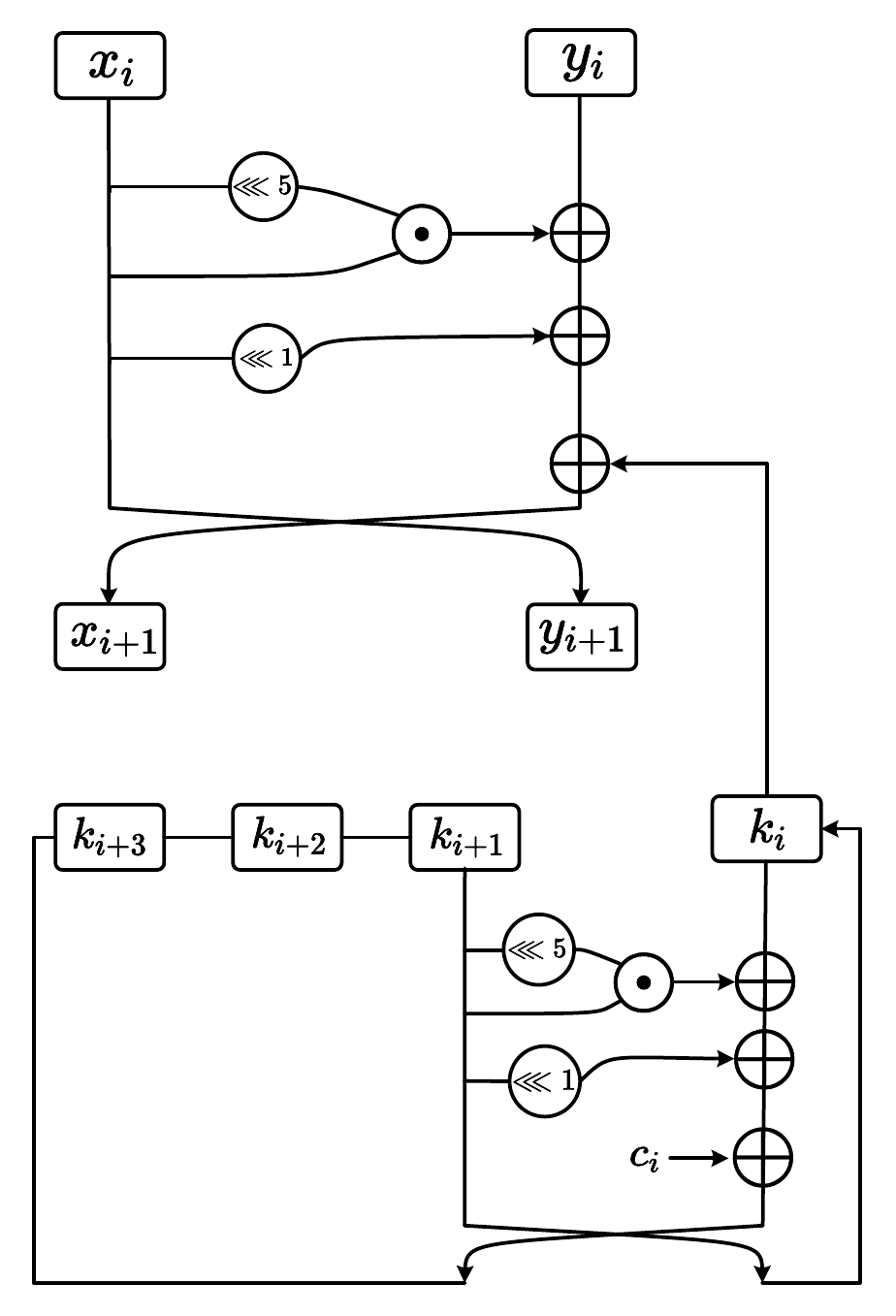}\label{fig:2simeck_cipher}}
        
        \caption{The round functions and key schedules of Simon and Simeck }
        \label{fig:2simeck_simon_cipher}
    \end{figure}

Simon~\cite{beaulieu2013simon} is a family of lightweight block ciphers published by the National Security Agency based on the Feistel structure. A member of the family is denoted by \text{Simon}$2n/mn$, where $n$ is the branch size, $2n$ is the block size, and $mn$ is the key length, $n\in \{16,24,32,48,64\}$, $m\in \{2,3,4\}$. The round function consists of cyclic rotation ($\lll$), bitwise AND ($\odot$), and bitwise XOR ($\oplus$). The key schedule is recursive and divided into 3 types based on the value of $m$. This paper only focuses on the key schedule when $m=4$ and its formula is defined as 
$k_{i}=k_{i-4}\oplus ((k_{i-3}\oplus (k_{i-1}\ggg 3))\oplus((k_{i-3}\oplus (k_{i-1}\ggg 3)))\ggg 1))\oplus c$, where $c_i$ is a round constant. 
In 2015, Yang et al.~\cite{yang2015simeck} proposed the Simeck family. They chose the different rotation offsets in round function, and reuse the round function as its key schedule which leads to better implementation in hareware than Simon. Because we devote attention to Simon32/64 and Simeck32/64, so we only list their round functions and key schedule in Fig.~\ref{fig:2simeck_simon_cipher}.

\subsection{Rotational-XOR Cryptanalysis}
Rotational-XOR (RX) cryptanalysis, originally proposed by Ashur and Liu~\cite{ashur2016rotational} at FSE 2016, is an improvement over rotational cryptanalysis aiming to address the problem that the rotational-invariant property of internal state cannot hold if round constants are injected into key schedule.

\begin{definition}[RX-difference~\cite{lu2022improvedsimon_like}] The \textit{RX-difference} of $x$ and $x'=(x \lll \lambda) \oplus \alpha$ is denoted by 
$$
\Delta_\lambda {(x,x')}=(x\lll \lambda)\oplus x'=\alpha
$$
where $\alpha \in \mathbb{F}^{n}_{2}$ is a constant and $\lambda$ is a rotational offset with $0<\lambda<n$, and $(x,x')$ is called an RX pair.
\end{definition}

In this paper, our target ciphers are Simon and Simeck, both of which are with Feistel structure. Also, we only focus the initial RX-difference with zero left branch and non-zero right branch. For the sake of convenience, we define this special initial RX-difference as \textit{half RX-difference} that is clarified as follows.

\begin{definition}[Half RX-difference]
    For a Feistel based block cipher with size of $2n$, the \textit{half RX-difference} $[\lambda, \Delta_{R}]$ is defined as the input RX-difference, in which the left (resp. right) branch $\Delta_L\in\mathbb{F}_2^n$ ($\Delta_R\in\mathbb{F}_2^n$) is inactive (resp. active) and the rotational offset is $\lambda$.
\end{definition}

\section{Finding Good RX-differences and Fundamental Data Formats}
\label{sec:3_0}	
 For constructing the better RX-neural distinguishers, it is of great significance to prepare better half RX-differences firstly. In this section, we will explain how to experimentally finding the better half RX-differences. First, we will introduce two new fundamental data formats. Next, we exhaust all the possible half RX-differences with Hamming weights of 1 and 2. Based on the half RX-differences and new data formats, we then construct datasets and train the RX-neural distinguishers. Finally, according to the accuracy of obtained RX-neural distinguishers, we determine the half RX-differences.
    
We first provide a detailed explanation of two data formats used in our experiments. The first one is inspired by Lu et al.'s data format proposed in~\cite{lu2022improved}. In particular, their data format is represented as $(\delta_{L}^{r},\delta_{R}^{r},C_L^{r},C_R^{r},{C'}_L^r,{C'}_R^r,\delta_{R}^{(r-1)},\underbar{$\delta$}_{R}^{(r-2)})$ where $\delta$ denotes the traditional difference. Specially, the $r$-round ciphertexts $(C^r,C'^r)$ are generated by a pair of plaintexts that satisfy a given difference, and $\underbar{$\delta$}_{R}^{(r-2)}$ that represents the left branch of $(r-2)$-round difference is derived by decrypting $(C^r,C'^r)$ for two rounds with the zero roundkeys. The novelty of this data format is that it makes use of the decrypted $(r-2)$-round information. Note that this data format is applied to differential-neural cryptanalysis. Therefore, to fit RX cryptanalysis, we adjust this data format by imposing rotation on $C^r$ and replacing traditional difference $\delta$ by RX-difference $\Delta$. Consequently, the first data format is $(\Delta _{L}^{r},\Delta _{R}^{r},C_L^{r}\lll \lambda,C_R^{r}\lll \lambda,{C'}_L^r,{C'}_R^r,\Delta _{R}^{(r-1)},\underbar{$\Delta$}_{R}^{(r-2)})$, where $\underbar{$\Delta$}_{R}^{(r-2)}$ is derived by decrypting $(C^r,C'^r)$ for two rounds with the zero roundkeys. We denote it by $\mathscr{D}_1$. The second one is based on Gohr's data format~\cite{gohr2019improving}, which is also proposed for differential-neural cryptanalysis and only consists of a pair of $r$-round ciphertexts. This data format is $(C^r_L,C^r_R,C'^r_L,C'^r_R)$. Similar to the adjustment on Lu et al.'s data format, we impose rotation on $C^r$ and get our second data format as $(C_L^{r}\lll \lambda, C_R^{r}\lll \lambda,{C'}_L^{r},{C'}_R^{r})$. We denote it by $\mathscr{D}_2$.

By investigating the previous good traditional RX distinguishers in~\cite{lu2022improvedsimon_like} as well as the RX-neural distinguishers in~\cite{palmierideep}, we found that the Hamming weight of initial RX-differences are generally small, in particular, they are 1 or 2. Therefore, we select all half RX-differences with the Hamming weight of 1 and 2 under the rotation offsets from 1 to 15 as candidates. In total, there are $15\times \left(\binom{16}{1}+\binom{16}{2}\right)=2040$ candidates of half RX-difference for Simon32/64 and Simeck32/64. For each candidate, we first randomly choose $2^{23}$ pairs of plaintexts satisfying the half RX-difference and encrypt them to generate the pairs of ciphertexts, respectively for 11-round Simon32/64 and 13-round Simeck32/64. Then we use the prepared data format $\mathscr{D}_1$ and $\mathscr{D}_2$ to construct the training datasets. Moreover, the \textbf{hyper-paramter} of our experiments\footnote{In this work, all experiments are conducted on a platform with four Intel (R) Xeon (R) E5-2698 v4 @ 2.20GHz CPUs, four Tesla V100 DGXS 32GB GPUs, and the Ubuntu 20.04 system.} is set as: 
\begin{itemize}
    \item the size of validation dataset is $2^{18}$;
    \item the size of batch is $2^{15}$;
    \item the number of iterations is 10;
    \item the number of learning rates is 10, which vary from $0.0001$ to $0.1$ with the gap of $0.0111$.
\end{itemize}
Utilizing the distributed training strategy of Keras\footnote{The Keras Documentation is available via the link \url{https://keras.io}.}, each iteration takes 25 (resp. 20) seconds in average for $\mathscr{D}_1$ (resp. $\mathscr{D}_2$), and training one RX-neural network need to take about four minutes. As a result, it took about 23 days to accomplish the whole $2040\times 2\times 2 = 8160$ experiments for Simon and Simeck. 

In general, if the accuracy of a neural distinguisher is more than 0.5, then we regard it as a valid distinguisher for mounting attacks. Taking an accuracy of 0.51 as the standard, we consequently retain 1496 and 151 valid RX-neural distinguishers\footnote{Our experiments indicate that $\mathscr{D}_1$ is better than $\mathscr{D}_2$.} for 11-round Simon32/64 and 13-round Simeck32/64, respectively. In particular, 

\begin{itemize}
\item As for the 11-round Simon32/64, a total of 1496 valid distinguishers are identified. When the Hamming weight of the RX-difference is fixed to 1, the number of valid distinguishers is 208. Among these, there are 143, 54 and 9 distinguishers with the accuracy distributed at 0.51-0.55, 0.55-0.60 and 0.60-0.65. The remaining two have the accuracy of 0.6550 and 0.6543, corresponding to the half RX-differences $[13, 0x4000]$ and $[3, 0x2]$, respectively. Surprisingly, when the Hamming weight is 2, the number of valid distinguishers is 1288. Additionally, there are a total of 6 distinguishers with an accuracy greater than 0.80, as shown in Table~\ref{best_11simon_13simeck}.  

\item As for the 13-round Simeck32/64, when the Hamming weight of RX-difference is fixed to 1, the distribution of valid distinguishers exhibits a relatively concentrated pattern. For rotation offsets $\{1,4,5,11,12,15\}$ and $\{2,3,6,10,13,14\}$, there are 6 and 5 distinguishers, respectively. The remaining offsets $\{7,8,9\}$ have no valid distinguisher. When the Hamming weight is 2, the number of valid distinguishers becomes more widely distributed. For rotation offsets $\{1,15\}$, there are 18 valid distinguishers. For offsets $\{11\}$ and $\{4,5,12\}$, there are 13 and 10 distinguishers, respectively. But for $\{2\}$, $\{14,6,10\}$, $\{3,13\}$, $\{7\}$, and $\{9\}$, there are only 6, 5, 3, 2, and 1 distinguishers, respectively. The offset $\{8\}$ has no valid distinguisher. Among these results, the best accuracy is closed to 0.7, with half RX-differences of $[1, 0x4]$ and $[15, 0x2]$.

\end{itemize}

For more details, please refer to Sect.~I of the online Supplemental Material\footnote{\url{https://maipdf.cn/file/dt68808759e74d4/pdf} or \url{https://gitee.com/anonymous_ieee/anonymous_ieee.git}}. By sorting these in descending order of accuracy, we here only list the top six candidates in~\ref{best_11simon_13simeck}. Also, we compare the accuracy of these RX-neural distinguishers with the previous results~\cite{palmierideep}, all of ours are better than or consistent to that of~\cite{palmierideep}, which demonstrates that our proposed data formats and the method to construct half RX-differences are indeed effective. It is worth noting that the results of~\cite{palmierideep} listed in Table~\ref{best_11simon_13simeck} are the optimized one by exploiting multi-ciphertext data format but ours do not. Hence, we are sure that the accuracy of our presented RX-neural distinguisher can be further improved, which will be detailed in the next section. 

\begin{table}[!h]
    \centering
    \renewcommand\tabcolsep{6.5pt}
	\caption{The six half RX-differences for 11-round Simon32/64 and 13-round Simeck32/64 RX-neural distinguishers using $\mathscr{D}_1$}
	\label{best_11simon_13simeck}
	\begin{tabular}{lclcc}
       \toprule
		Simon  & Accuracy  & Simeck  & Accuracy & Ref. \\
		\midrule
            $[3, 0x2]$ & 0.5445  &$[1, 0x2]$ & 0.7057 & ~\cite{palmierideep}\\ 
            $[15, 0x3]$ & 0.9215   &$[1, 0x4]$ & 0.6990 & Ours\\ 
		$[1, 0x6]$ & 0.9203  &$[15, 0x2]$ & 0.6989 & Ours\\ 
		$[12, 0x2002]$ & 0.8823  & $[15, 0x3]$ & 0.6202 &Ours\\ 
		$[4, 0x22]$ & 0.8802  &$[1, 0x6]$ & 0.6212 & Ours\\ 
		$[3, 0x12]$ & 0.8115 &$[6, 0x82]$ & 0.6184 & Ours\\ 
		$[13, 0x4002]$ & 0.8097  &$[10, 0x802]$ & 0.6168 & Ours\\
        \bottomrule
	\end{tabular}
\end{table}

\section{Exploring Better RX-Neural Distinguishers}
\label{sec:4_0}
As illustrated in the last section, we find some good half RX-differences that can derive the high-accuracy RX-neural distinguishers even if we do not exploit the multi-ciphertext data formats. To obtain higher-accuracy or longer-round ones, it is of great significance to add multi-ciphertext into the components of the two data formats. On the one hand, according to the previous studies~\cite{lu2022improved,chen2023new,hou2021improve,benamira2021deeper,gohr2022assessment}, we know that the more the number of multi-ciphertext components contained in a data format, the higher the accuracy of the trained neural distinguisher. On the another hand, the number of components of a data format has a directly influence on the required memory space (it is GPU memory in general) for training the neural network. If the required memory approaches or exceeds the total memory of the machine, the training process will be terminated. Therefore, we must strike a balance between the accuracy of the neural distinguishers and the memory requirements of the training process, under a limited computational resources. In other words, we need to control the number of total components as much as possible while adding the multi-ciphertext into data formats. 

Regard $\mathscr{D}_1$ as the basic data format, we first remove some redundant components, which will be detailed in Sect.~\ref{sec:4_1}. Then in Sect.~\ref{sec:4_2}, we add multi-ciphertext into the simplified data formats, and train distinguishers using multi-ciphertext data formats. To explore RX-neural distinguishers that cover more rounds, we in Sect.~\ref{sec:4_3} exploit the staged training method to extend the trained ones from Sect.~\ref{sec:4_2}.

\subsection{Removing Redundant Components from Data Format $\mathscr{D}_1$}\label{sec:4_1}
To remove components from $\mathscr{D}_1$, we need to detect the impact of ciphertexts on the accuracy of neural disinguishers and then determine the redundant components. Here we utilize the \textit{bit sensitivity test} (BST) method proposed in~\cite{chen2020neural}. The main idea of this method is to randomly modify the value of a fixed bit position in the ciphertexts of each sample in the validation dataset by XORing with a random mask, and then analyze the impact on the accuracy of neural distinguisher. We apply BST method to the six neural distinguishers in Table~\ref{best_11simon_13simeck} for Simon32/64 and Simeck32/64. Consequently, we conclude that $\Delta_L^r$, $C^r_L\lll\lambda$ and $C'^r_L$ are the redundant components of $\mathscr{D}_1$. Next, we will give a detailed illustration about the process of BST.

As an example, we focus the 11-round neural distinguisher for Simon32/64 with the half RX-difference of [15, 0x3] and the accuracy of 0.9215. To detect the redundant components of $\mathscr{D}_1$, we first randomly choose $2^{18}$ pairs of plaintexts satisfying the half RX-difference of [15, 0x3] to generate an original set that contains $2^{18}$ pairs of ciphertexts. Taking BST on the least significant bit (the bit position is 0) as an instance, we randomly choose $2^{18}$ 32-bit masks, where the value of the bit position 0 varies, and the values of other bit positions are all 0s. Then, we XOR these masks with the elements in the original pair-of-ciphertexts set one by one to generate a modified set. Here, we define three XORing types according to the objective pair of ciphertexts $(C,C')$ as XORing with the single ciphertext ($C$ or $C'$) or pair of ciphertexts might lead to different effects. The first type, denoted as \texttt{TYPE1}, involves XORing the mask only with $C$, which corresponds to $(C^r_L,C^r_R)$ of the data format $\mathscr{D}_1$. On the contrary, the second type, denoted as \texttt{TYPE2}, XORs the mask only with $C'$, corresponding to $(C'^r_L,C'^r_R)$ of $\mathscr{D}_1$. The third type, denoted as \texttt{TYPE3}, involves XORing the mask with both $C$ and $C'$. 
    For each of these types, a modified pair-of-ciphertexts set is generated by applying the corresponding XOR operation to the original one. Based on the data format $\mathscr{D}_1$, three types of validation dataset can be generated from the modified pair-of-ciphertexts sets. Moreover, the validation dataset of each type is put into the 11-round neural distinguisher of Simon32/64 to obtain a verified accuracy. Finally, the difference between the original accuracy (i.e., 0.9215) and the verified accuracy can be calculated, which is regarded as the \textit{bit sensitivity}. By the way, the smaller the value of bit sensitivity, the smaller the influence of the corresponding bit on the accuracy.

\begin{figure*}[!b]
	\centering
        \subfloat[Training using {$[15, 0x3]$}]{\includegraphics[width=2.4in]{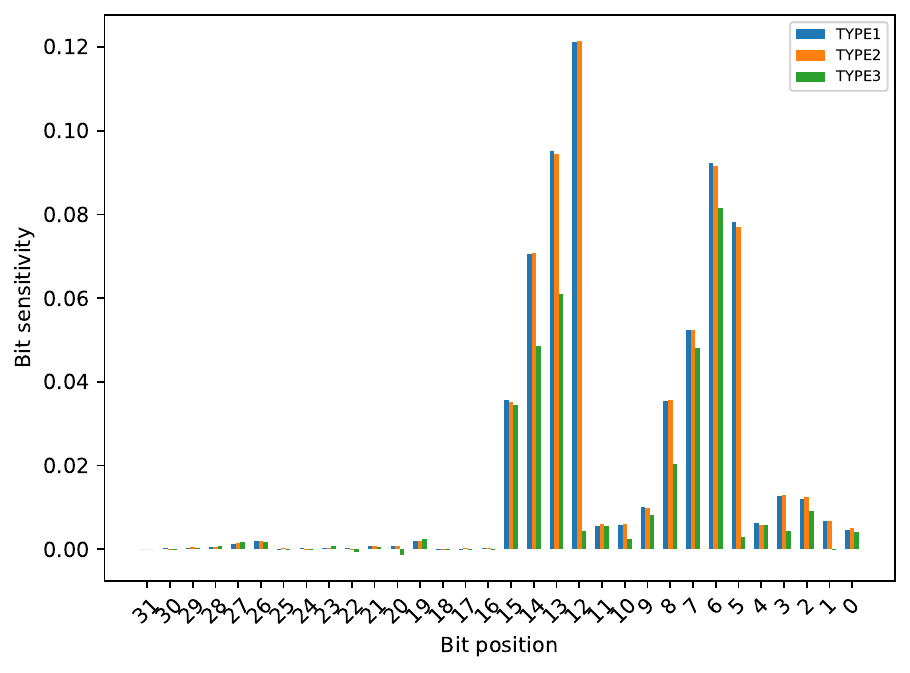}\label{fig:simon32_data14_15_3_11r_bit_sensitivity}}
        \hspace{0.8cm}
        \subfloat[Training using {$[1, 0x4]$}]{\includegraphics[width=2.4in]{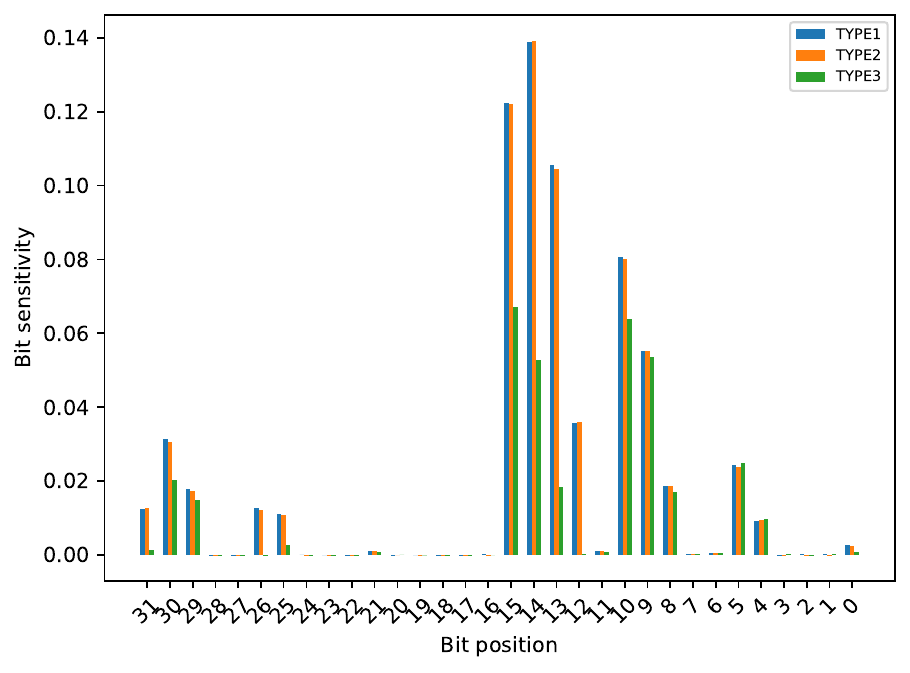}\label{fig:simeck32_data14_1_4_13r_bit_sensitivity}}
        \caption{The bit sensitivity for 11-round Simon32/64 (left) and 13-round Simeck32/64 (right), both of which use the data format $\mathscr{D}_1$. For more bit sensitivity test, refer to Sect.~II of Supplemental Material.
        }
	\label{fig:bst1_simon}
\end{figure*}
    
    By repeating the above process for the remaining 31 bit positions, we can obtain the sensitivities for each ciphertext bit across the three types. For a more intuitive comparison, we present the results in Fig.~\ref{fig:bst1_simon}, from which it can be observed that all bits on the left branch of the ciphertext (note that 31 represents the most significant bit position) have the negligible influcence on the accuracy of the neural distinguisher, as the bit sensitivities are close to or equal to 0s. In other words, the components related to the left branch of ciphertext in $\mathscr{D}_1$, i.e. $\Delta_L^r$, $C^r_L\lll\lambda$ and $C'^r_L$ are redundant. Similarly, we apply BST method to the 13-round neural disitinguisher of Simeck32/64 with the half RX-difference [1, 0x4] and the accuracy of 0.7 and exhibit the bit sensitivities in Fig.~\ref{fig:simeck32_data14_1_4_13r_bit_sensitivity}. Note that there are only five bits on the left branch of cipher can effect the accuracy. Besides, the corresponding bit sensitivities are quite small. Therefore, we can also regard $\Delta_L^r$, $C^r_L\lll\lambda$ and $C'^r_L$ as the redundant components of $\mathscr{D}_1$. 


Thus, by removing the redundant components from $\mathscr{D}_1$ as mentioned above, we obtained the simplified data format 
$(\Delta _{R}^{r},C_L^{r}\lll \lambda,C_{R}^{r},\Delta _{R}^{(r-1)},\underbar{$\Delta$}_{R}^{(r-2)})$, denoted by $\mathscr{D}_3$. Moreover, we wonder whether the data format  $\mathscr{D}_3$ could be further simplified. Note that $\mathscr{D}_3$ contains two types of components: ciphertexts and RX-differences. We try to remove the two types of components from $\mathscr{D}_3$ to respectively obtain the simplified data formats $\mathscr{D}_4$ and $\mathscr{D}_5$. As shown in Table~\ref{data_component}, the accuracy of distinguisher trained using $\mathscr{D}_4$ has a significant decrease compared to $\mathscr{D}_3$ (from 0.9221 to 0.5013). Nevertheless, $\mathscr{D}_5$ has the nearly same accuracy as $\mathscr{D}_3$. Therefore, the ciphertexts i.e., $C_R^r\lll \lambda$ and $C'^r_R$, are the redundant components of $\mathscr{D}_3$. Similarly, regard $\mathscr{D}_5$ as a basic data format, we try to remove each one of the three components respectively to obtain $\mathscr{D}_6$, $\mathscr{D}_7$, and $\mathscr{D}_8$, and the corresponding 
trained accuracies are 0.7641, 0.9211, and 0.5912 as shown in Table~\ref{data_component}. This indicates that $\Delta_{R}^{r}$ and $\Delta_{R}^{(r-1)}$ are the necessary components, while $\Delta_{R}^{(r-2)}$ is a redundant component in $\mathscr{D}_5$.

 \begin{table*}[!h]
		\begin{center}
			\caption{Data formats (written as DF in short) and the corresponding accuracy of RX-neural distinguishers for 11-round Simon32/64. Note the half RX-difference is [15, 0x3], thus the value of $\lambda$ is 15}
           \renewcommand\tabcolsep{6.5pt}
			\begin{tabular}{clc}
				\toprule
				DF & Specification &Accuracy \\ \midrule
    
				$\mathscr{D}_{1}$&$(\Delta _L^{r},\Delta _R^{r},C_L^r\lll \lambda ,C_R^r\lll \lambda ,{C'}_L^r,{C'}_R^r,\Delta _R^{(r-1)},\underbar{$\Delta$} _R^{(r-2)})$  & \textbf{0.9215} \\
    
                $\mathscr{D}_{3}$&$(\Delta _R^{r},C_R^r\lll \lambda ,{C'}_R^r,\Delta _R^{(r-1)},\underbar{$\Delta$} _R^{(r-2)})$ & \textbf{0.9221} \\
    
                $\mathscr{D}_{4}$&$(C_R^r\lll \lambda ,{C'}_R^r)$  & 0.5013 \\
                
                $\mathscr{D}_{5}$&$(\Delta _R^{r},\Delta _R^{(r-1)},\underbar{$\Delta$} _R^{(r-2)})$ & \textbf{0.9204} \\

                $\mathscr{D}_{6}$&$(\Delta _R^{r-1},\underbar{$\Delta$} _R^{(r-2)})$ & 0.7641   \\

                $\mathscr{D}_{7}$&$(\Delta _R^{r},\Delta _R^{(r-1)})$   & \textbf{0.9211}  \\

                $\mathscr{D}_{8}$&$(\Delta _R^{r},\underbar{$\Delta$} _R^{(r-2)})$   & 0.5912  \\
                
                \bottomrule
			\end{tabular}
			\label{data_component}
		\end{center}
	\end{table*}

\subsection{Training Distinguishers using Multi-ciphertext Data Formats}
\label{sec:4_2}
As mentioned earlier, the more multi-ciphertext a data format contains, the higher the corresponding accuracy will be. Due to the limitation of GPU memory of our platform, the maximal number of multi-ciphertext that can be accommodated in $\mathscr{D}_5$ and $\mathscr{D}_7$, determined by a series of experimentally tests, are 28 and 36 respectively. Consequently, we derive the longest extended data formats from $\mathscr{D}_5$ and $\mathscr{D}_7$, denoted by $\mathscr{D}_5^{28}$ and $\mathscr{D}_7^{36}$, which adopt 28 and 36 pairs of ciphertexts. Applying the two data formats as well as $\mathscr{D}_5$ to 11-round Simon32/64 under the six half RX-differences (listed in Table~\ref{best_11simon_13simeck}), we have the trained results as presented in Table~\ref{table:simon_acc}. 
\begin{table}[!htp]
		\begin{center}
			\caption{Comparison on the accuracy of trained RX-neural distinguishers for Simon32/64 using different data format. Note DF means data format.The underlined items are used for key-recovery attacks in Sect.~\ref{sec:5_0}}
              \renewcommand\tabcolsep{1.7pt}
			\label{table:simon_acc}
			\begin{tabular}{cccccccc}
				\toprule
			    \#R & DF & $[15, 0x3]$ & $[1, 0x6]$ & $[12, 0x2002]$ & $[4, 0x22]$ & $[13, 0x4002]$ & $[3, 0x12]$\\ \midrule
		    \multirow{3}{*}{11r} & $\mathscr{D}_5$ & 0.9204 & 0.9204 & 0.8869 & 0.8905 & 0.8165 & 0.8170 \\ 
			& $\mathscr{D}_5^{28}$ & 0.9999 & 0.9999 & 0.9997 & 0.9996 & 0.9999 & 0.9997 \\ 
                & $\mathscr{D}_7^{36}$ & 1.0000 & 1.0000 & 0.9999 & 0.9998 & 0.9999 & 0.9999 \\ 
                \midrule
                \multirow{3}{*}{12r} & $\mathscr{D}_5$ & 0.5921 & 0.5946 & 0.6482 & 0.6437 & 0.5798 & 0.5750 \\ 
			& $\mathscr{D}_5^{28}$ & 0.9236 & 0.9336 & 0.9744 & \underline{\textbf{0.9713}} & 0.8991 & 0.8827 \\ 
                & $\mathscr{D}_7^{36}$ & 0.9176 & 0.9065 & 0.9821 & 0.9832 & 0.9075 & 0.9063 \\ 
                
                \midrule
                
                \multirow{3}{*}{13r} & $\mathscr{D}_5$ & 0.5100 & 0.5115 & 0.5249 & 0.5227 & 0.5091 & 0.5080 \\ 
			& $\mathscr{D}_5^{28}$ & 0.6300 & 0.6301 & \textbf{0.7120} & \underline{\textbf{0.7105}} & 0.6837 & 0.6848 \\ 
                & $\mathscr{D}_7^{36}$ & 0.6086 & 0.5962 & 0.6893 & 0.6872 & 0.5889 & 0.6062 \\

                \midrule

                \multirow{2}{*}{14r} & $\mathscr{D}_5^{28}$ & 0.5006 & 0.5008 & 0.5009 & 0.5005 & 0.5004 & 0.5000 \\ 
                & $\mathscr{D}_7^{36}$ & 0.5002 & 0.5001 & 0.5002 & 0.5003 & 0.5006 & 0.5004 \\
                \bottomrule
			\end{tabular}
		\end{center}
	\end{table}
 
It can be seen that the accuracy corresponding to $\mathscr{D}_5$ under each half RX-difference is actually identical to $\mathscr{D}_1$ (see in Table~\ref{best_11simon_13simeck}). Additionally, either $\mathscr{D}_5^{28}$ or $\mathscr{D}_7^{36}$ can achieve an exceptionally high accuracy, nearly reaching $100\%$. Apparently, this appears that it is potential to leverage $\mathscr{D}_5^{28}$ or $\mathscr{D}_7^{36}$ to train neural distinguishers covering longer rounds. Thus, we train the neural distinguishers for 12 to 14 rounds of Simon32/64, and give the corresponding accuracy in Table~\ref{table:simon_acc}. The accuracies of 14-round trained distinguishers shown in this table are more than 0.5, but in fact, they are not distinguishable. This is because the accuracy recorded in the table is the maximum value among all epochs. Besides, we conducted prediction experiments on the trained distinguishers, and the actual accuracies are unstable at 0.5 within the allowable range of statistical error, indicating that these distinguishers are invalid. As a consequence, the longest valid RX-neural distinguisher for Simon32/64 trained using multi-ciphertext can reach up to 13 rounds, with the best accuracy of 0.7120. Compared to the previouly best RX-neural distinguisher presented in~\cite{palmierideep}, our result not only extends the round (from 11 to 13) but also improves the accuracy (from 0.5445 to 0.7120). 

Similarly for Simeck32/64 (from 13 to 16 rounds), we trained a series of RX-neural distinguishers using the three data formats ($\mathscr{D}_5$, $\mathscr{D}_5^{28}$ and $\mathscr{D}_7^{36}$) and the six half RX-differences. We list the accuracy in Table~\ref{table:simeck_acc}. It can be seen that training with the two multi-ciphertext data formats can actually derive the 16-round distinguishers with a stable accuracy more than 0.51. Note that the currently best RX-neural distinguisher, presented in~\cite{palmierideep}, covers 15 rounds and has the accuracy of 0.5475. Our trained results improve the accuracy of 15-round one to 0.6022, notably extend the number of round to 16.
\begin{table}[!h]
		\begin{center}
			\caption{Comparison on the accuracy of trained RX-neural distinguishers for Simeck32/64 using different data format. Note DF means data format. The underlined items are used for key-recovery attacks in Sect.~\ref{sec:6_0}}
              \renewcommand\tabcolsep{2.5pt}
			\label{table:simeck_acc}
			\begin{tabular}{cccccccc}
                \toprule
                \#R & DF & $[1, 0x4]$ & $[15, 0x2]$ & $[15, 0x3]$ & $[1, 0x6]$ & $[6, 0x82]$ & $[10, 0x802]$\\ \midrule
		    \multirow{3}{*}{13r} & $\mathscr{D}_5$ & 0.6988 & 0.6997 & 0.6202 & 0.6214 & 0.6215 & 0.6198 \\ 
			& $\mathscr{D}_5^{28}$ & 0.9962 & 0.9999 & 0.9997 & 0.9997 & 0.9996 & 0.9997 \\ 
                & $\mathscr{D}_7^{36}$ & 0.9996 & 0.9995 & 0.9970 & 0.9967 & 0.9966 & 0.9968 \\ 
                
                \midrule
                
                \multirow{3}{*}{14r} & $\mathscr{D}_5$ & 0.5774 & 0.5789 & 0.5250 & 0.5262 & 0.5009 & 0.5014 \\ 
			& $\mathscr{D}_5^{28}$ & 0.9204 & 0.9136 & 0.7317 & 0.7326 & 0.5000 & 0.5000 \\ 
                & $\mathscr{D}_7^{36}$ & \underline{\textbf{0.8982}} & 0.8982 & 0.7095 & 0.7097 & 0.5019 & 0.5013 \\ 
                
                \midrule
                
                \multirow{3}{*}{15r} & $\mathscr{D}_5$ & 0.5120 & 0.5136 & 0.5026 & 0.5016 & 0.5023 & 0.5016 \\ 
			& $\mathscr{D}_5^{28}$ & 0.5905 & 0.5923 & 0.5191 & 0.5189 & 0.5000 & 0.5000 \\ 
                & $\mathscr{D}_7^{36}$ & \underline{\textbf{0.6042}} & \textbf{0.6022} & 0.5206 & 0.5250 & 0.5025 & 0.5000 \\

                \midrule

                \multirow{2}{*}{16r} & $\mathscr{D}_5^{28}$ & 0.5066 & \textbf{0.5127} & 0.5004 & 0.5006 & 0.5005 & 0.5008 \\ 
                & $\mathscr{D}_7^{36}$ & \underline{\textbf{0.5130}} & 0.5057 & 0.5007 & 0.5008 & 0.5010 & 0.5009 \\
                \bottomrule
			\end{tabular}
		\end{center}
	\end{table}

\subsection{Extending Distinguishers with Staged Training Method}\label{sec:4_3}
By leveraging the multi-ciphertext data format, either the round or the accuracy of RX-neural distinguishers for Simon32/64 and Simeck32/64 are indeed improved. However, if we only use the basic training scheme when training the neural distinguisher, it is difficult to break through 14 and 16 rounds for Simon32/64 and Simeck32/64, respectively. Therefore, we apply the \textit{staged training} scheme to enhancing the training process and hope to explore valid neural disinguishers covering more rounds. The concept of staged training was originally introduced by Gohr~\cite{gohr2019improving} at CRYPTO 2019 based on the idea of reinforcement learning, and later has become a crucial approach to enhance the neural-based distinguishers~\cite{bao2022enhancing,lu2022improved}. It turns an already trained ($r-1$)-round distinguisher into an $r$-round distinguisher in several stages. In addition, the accuracy of derived $r$-round distinguisher relies on that of the $(r-1)$-round one. 

The application of staged training to Simon32/64 can be divided into three stages. In the first stage, we use
the 13-round disinguisher, which was already trained using $\mathscr{D}_5^{28}$ and [4, 0x22] (resp. [12, 0x2002]) with accuracy $0.7105$ (resp. $0.7120$), to recognize the 12-round encryption with the same data format and half RX-difference, i.e., $\mathscr{D}_5^{28}$ and [4, 0x22] (resp. [12, 0x2002]).
This stage was done on $2^{23}$ training samples and $2^{20}$ testing samples for 10 epochs. Note that all the samples are derived from the 12-round multi-ciphertext. The batch size is $2^{18}$ and learning rate is ranged from 0.0001 to 0.00001. In the second stage, we adopt the updated network from the first stage to recognize the 14-round dataset using the same data format and half RX-difference. In this stage, the hyper-parameter remain consistent with those from the first stage, except that the samples are generated by the 14-round multi-ciphertext and the number of training samples is increased to $2^{24}$. In the third stage, we fresh the $2^{24}$ training samples and fed them to the network from the second stage. The number of epoch is set to 40 or more, and the learning rate is fixed to 0.00001. Finally, by this staged training strategy, we retain two valid RX-neural distinguishers covering 14 rounds with the respective accuracies of 0.5190 and 0.5240, as listed in Table~\ref{table:staged_result}.

However, for Simeck32/64, the staged training method successfully applied to Simon32/64 failed. No valid 17-round RX-neural distinguisher could be explored. Thus, we skipped the first stage and straightforwardly used the two 16-round distinguishers with the accuracies of 0.5130 and 0.5057 to recognize the corresponding 17-round datasets in the second stage. After using a lower learning rate of $10^{-5}$ in the third stage, we obtained a 17-round distinguisher with an accuracy of 0.5040, correspongding to the half RX-difference [1, 0x4] (shown in Table~\ref{table:staged_result}). This accuracy seems a little weak, but in fact, it is very stable and effective in the model prediction experiments.
\begin{table}[!h]
    \begin{center}
        \caption{Enhanced RX-neural distinguishers for Simon32/64 and Simeck32/64 by staged training. DF: data format. HRXD: half RX-difference.The underlined items are used for attacks.The underlined items are used for key-recovery attacks in Sect.~\ref{sec:5_0} and~\ref{sec:6_0}}
        \renewcommand\tabcolsep{2pt}
        \begin{tabular}{lclccccc}
            \toprule
            Cipher & DF & HRXD & \#R & Accuracy & TPR & TNR & Validity \\ \midrule

             \multirow{2}{*}{Simon32/64} & \multirow{2}{*}{$\mathscr{D}_{5}^{28}$} & $[12, 0x2002]$ & {14} & 0.5190 & 0.5176 & 0.5204 & \ding{51}\\
            &  & $[4, 0x22]$  & {14} & \underline{\textbf{0.5241}} & 0.5634 & 0.4848 & \ding{51}\\
    
            \cmidrule{1-8}

            \multirow{2}{*}{Simeck32/64} & \multirow{2}{*}{$\mathscr{D}_{7}^{36}$} &  $[1, 0x4]$  & {17} & \underline{\textbf{0.5040}} & 0.5950 & 0.4130 & \ding{51}\\ 
            &  & $[15, 0x2]$ & {17} & 0.5000 & - & - & \ding{55} \\ 
            \bottomrule
        \end{tabular}
        \label{table:staged_result}
    \end{center}
\end{table}

\section{Key-recovery Attacks on Simon32/64}
\label{sec:5_0}
In this section, we will give the details of recovering key bits of the round-reduced Simon32/64 based on the trained RX-neural distinguishers. Different from the attacks based on conventional distinguishers, the neural-based key-recovery attacks need to use the Bayesian key-recovery strategy (BKS) to guess key bits. Previous works also combine with the techinque used in conventional distinguisher based key-recovery attacks, like extending round forward or backward on distinguishers, to mount attacks covering more rounds~\cite{gohr2019improving,bao2022enhancing,zhang2022improving,zhang2023improved}. In this paper, we consider to extend the RX-neural distinguisher backward by only one round, and straightforwardly use BKS to achieve practical key-recovery attacks. 
\begin{adjustwidth}{0.4cm}{0cm}  

    \begin{minipage}{0.9\columnwidth}  
    \begin{algorithm}[H]
		\renewcommand{\thealgocf}{3}     
		\SetAlgoLined
		\KwIn{Ciphertext structure $C=C_0,...,C_{m\times k-1}$, RX-neural distinguisher $\mathcal{ND}$, set $S:=\{rk_0,...,rk_{n-1}\}$ composed of $n$ candidates of guessed subkey, number of key search iteration $l$.}
	    \KwOut{A List that contains $l\times n$ guessed candidates of subkey.}
		$S := \{k_0, k_1, \ldots, k_{n-1}\} \leftarrow$ choose at random without replacement from the set of all subkey candidates.

		$L \leftarrow \{\}$;
  
		\For{$j \in \{0, 1, \ldots, m - 1\}$}
		{
			$P_{i,k} \leftarrow Decrypt(C_i, k)$ for all $i \in \{0, 1, \ldots, m - 1\}, k \in S$.

			$v_{i,k} \leftarrow N(P_{i,k})$ for all $i, k$

			$w_{i,k} \leftarrow \log_2(v_{i,k}/(1 - v_{i,k}))$ for all $i \in \{0, \ldots, m - 1\}, k \in S$

			$w_k \leftarrow \sum_{i=1}^{n} v_{i,k}$ for all $k \in S$

			$L \leftarrow L \cup \{(k, w_k) \text{ for } k \in S\}$

			$m_k \leftarrow \sum_{i=0}^{n-1} v_{i,k}/n$ for $k \in \{k_0, \ldots, k_{n-1}\}$

			$\lambda_k \leftarrow \sum_{i=0}^{n-1} (m_{ki} - \mu_{ki \oplus k})^2/\sigma_{ki \oplus k}^2$ for $k \in \{0, 1, \ldots, 2^{16} - 1\}$;

			$S \leftarrow argsort_k(\lambda)[0 : n - 1]$
		}
		return $L$
  
		\caption{BayesianKeySearch: find a list of subkey candidates}
       \label{algori:bayesian}
	\end{algorithm}
    \end{minipage}
\end{adjustwidth}
\vspace{5pt}

We first describe the framework of BKS, which is used to recover $(r+1)$-round subkeys (we denote it by $rk^{(r+1)}$) for Simon32/64 based on an $r$-round RX-neural disinguisher trained using $k$-multi-ciphertext data format and a half RX-difference $[\lambda, \Delta_R]$, as follows:
\begin{enumerate}
    \item Randomly choose $m\times k$ pairs of plaintexts satisfying the half RX-difference $[\lambda,\Delta_R]$, query for the corresponding $(r+1)$-round pairs of ciphertexts. A ciphertext structure composed of $m\times k$ elements is generated.
    \vspace{4pt}
        \item Select $n$ ($n< 2^{16}$) guessed candidates of $rk^{(r+1)}$ (there are in total $2^{16}$ candidates as the subkey is size of 16), then use BayesianKeySearch algorithm, as depicted in Algorithm~\ref{algori:bayesian}, to generate $l\times n$ guessed candidates for $rk^{(r+1)}$ where $l$ is the number of key search iteration. Note that each candidate has a score.
        \vspace{4pt}
        \item Set the filtering threshold for $rk^{(r+1)}$ to $c_1$, retain the candidates whose scores are greater than $c_1$. If there is no candidate left, then go back the step 2 and re-select $n$ guessed candidates for $rk^{(r+1)}$. Otherwise, go to the step 4.
        \vspace{4pt}
        \item For each of the $n_1$ left candidates of $rk^{(r+1)}$, generate $m\times n$ guessed candidates for $rk_r$ like doing in the step 2. There are in total $m\times n\times n_1$ guessed candidate pairs of $(rk^{r},rk^{(r+1)})$, each one corresponds to a score.
        \vspace{4pt}
        
        \item Set the filtering threshold for $(rk^{r},rk^{(r+1)})$ to $c_2$, retain the pairs whose scores are greater than $c_2$. If there are $n_2$ candidates left, rank them by their scores and the pair corresponding to the highest score is regarded as the correct value of $(rk^{r},rk^{(r+1)})$. Otherwise, go back to the step 2 and repeat the process till a pair of $(rk^r,rk^{(r+1)})$ is retained.
\end{enumerate}
    
In the above mentioned $(r+1)$-round key-recovery framework based on BKS, BayesianKeySearch plays an important role. The values of $\mu$ and $\sigma$ in Line 9 of Algorithm~\ref{algori:bayesian} depend on the \textit{wrong key response} (WKR) of the targeting subkey. As the core step of BKS, the concept of WKR was introduced by Gohr at CRYPTO 2019, to optimize the candidates of guessed subkey such that the candidates that we generate cover the correct subkey with a high probability. One can refer to~\cite{gohr2019improving} for more detail, we will omit the WKR process and only exhibit the WKR profile in the concrete attack on Simon32/64. It is worth noting that the success rate and runtime of our attacks depend on the values of $c_1$ and $c_2$. In particular, if $c_1$ and $c_2$ is too large, the success rate will be high but the attack will take much more time. On the contrary, if $c_1$ and $c_2$ are small, the runtime will be less but the success rate will decrease. Therefore, to ensure that key-recovery attacks can be achieved
in a reasonable time with a high success rate, we need to set $c_1$ and $c_2$ to the appropriate values by conducting a pre-attack process before starting attacks. The process involves setting the correct key of the last 2 rounds as the initial guessed candidate, then observing the pre-attack score and finally determining threshold ranges through repeated experiments. 
\begin{figure*}[b]
        \centering
        \subfloat[For 14-round subkey]{\includegraphics[width=2.5in]{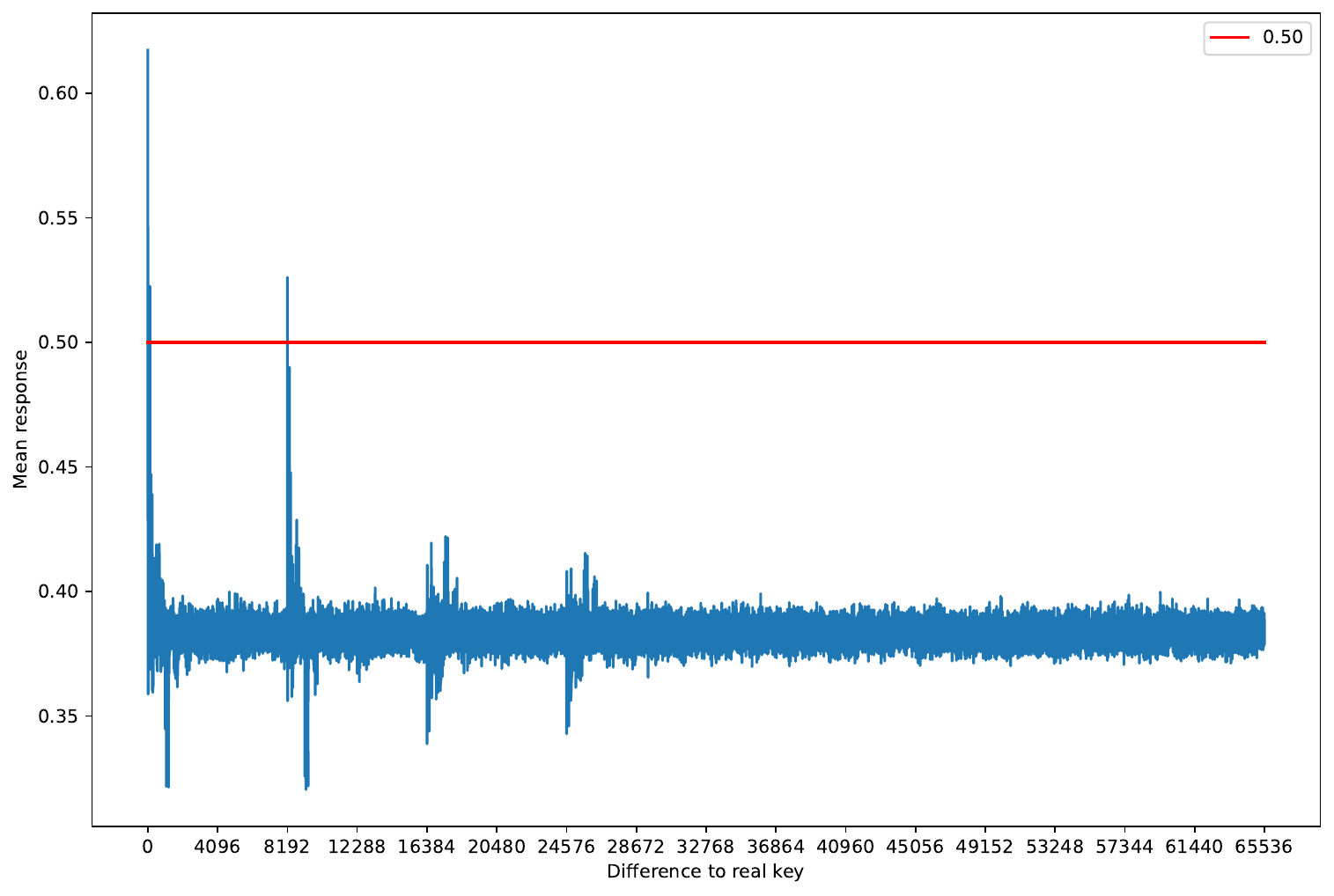}
            \label{fig:simon32_mean_data25_14r_pairs28}
        }
        \hspace{2cm}
        \subfloat[For 15-round subkey]{\includegraphics[width=2.5in]{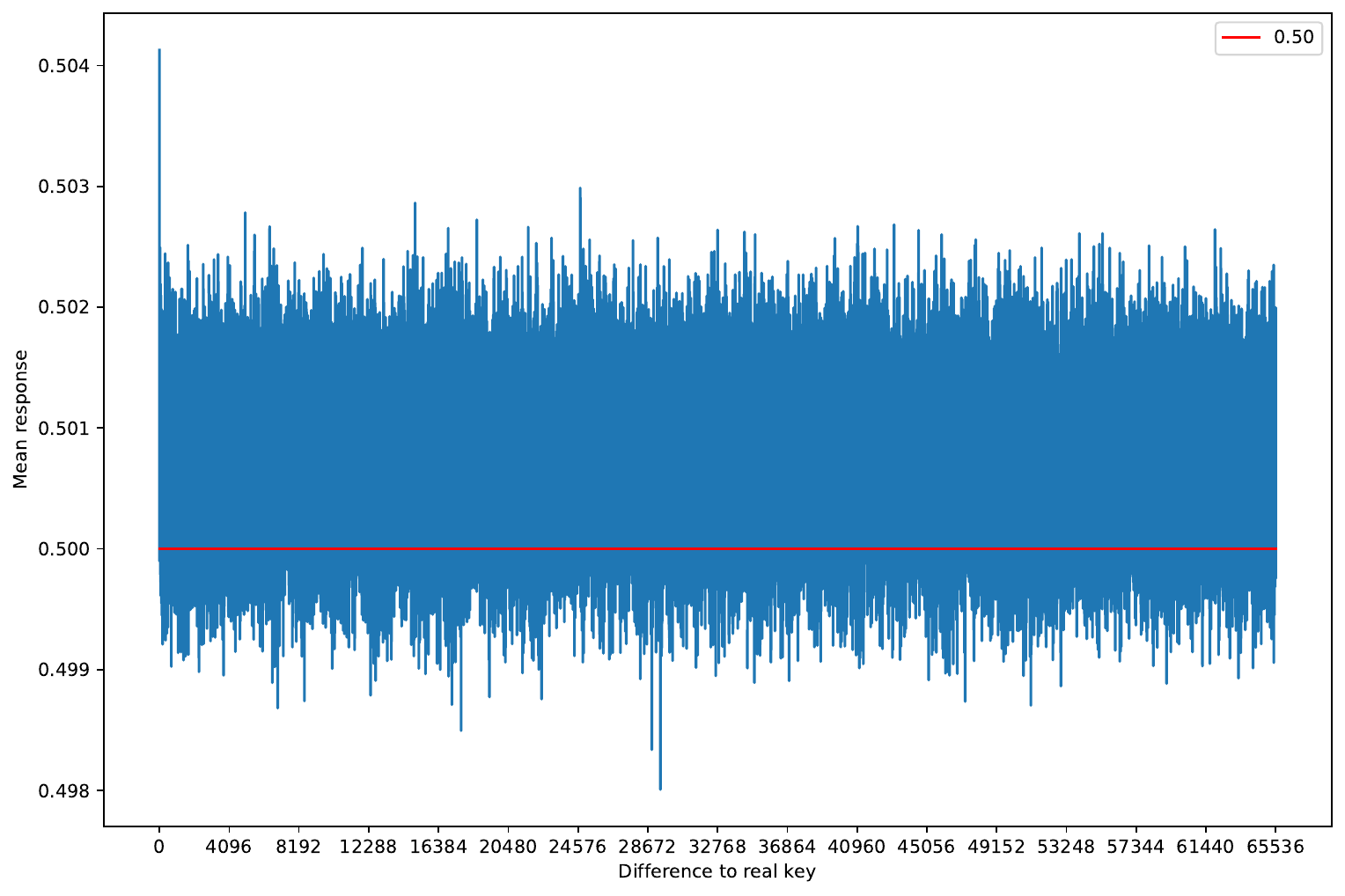}
        		\label{fig:simon32_mean_data25_15r_pairs28}
        }
        \caption{The WKS profiles for 14- and 15-round Simon32/64. For more WKS for other rounds, one can refer to Sect.~III of Supplemental Material}
        \label{fig:wrong_simon}
\end{figure*}

\vspace{5pt}
\noindent\textbf{Key-recovery attacks on 14- and 15-round Simon32/64.} Based on the explored RX-neural distinguishers and the key-recovery framework, we can straightforwardly achieve a 14- and 15-round key-recovery attacks on Simon32/64. Note that in the attacking framework, we expect to find a pair of guessed subkeys $(rk^r, rk^{(r+1)})$ as the correct one by repeating the process from step 2. In theory, it can actually return a pair as long as repeating for plenty of times. But it is not easy to evaluate the runtime in practice. Therefore, we need to set an upper bound on the repeating times when applying the framework to practical attacks, we denote it by $t$. If there is still no surviving candidate after $t$ times of repeating, then we just regard the candidate pair, which has the highest score among all $t$ times of repeating, as the correct subkeys.

\begin{itemize}
    \item To mount 14-round attack, we choose the 13-round RX-neural distinguisher with accuracy 0.7105, which is trained using half RX-difference $[4, 0x22]$ and data format {$\mathscr{D}_{5}^{28}$}. Under the same half RX-difference and data format, the accuracy of 12-round one is 0.9713. The attack parameters are set as:
\begin{center}
\renewcommand\tabcolsep{8pt}
\renewcommand{\arraystretch}{1.5} 
\begin{tabular}{llll}
                    $m = 2^{10}$ & \hspace{0.1cm} $k=28$ & \hspace{0.1cm} $c_{1} = 500$ & \hspace{0.5cm} $c_{2} = 1000$ \\
                    $t = 2^{10}$ & \hspace{0.1cm} $l = 6$ & \hspace{0.1cm} $n = 32$. & 
                \end{tabular}
            \end{center}
The WKR profile for 14-round subkey is exhibited in Fig.~\ref{fig:simon32_mean_data25_14r_pairs28}. The results demonstrate that the mean response of correct subkey is significantly different from the wrong ones. Notably, the correct subkey displays a mean response value exceeding 0.61, while the majority of wrong candidates remain below the 0.40 threshold. Namely, the recovered one possesses a high probability of matching the correct subkey. In fact, based on the high-accuracy RX-neural distinguisher, we can obtain a recovered pair of $(rk^{13}, rk^{14})$ within four minutes using the key-recovery framework. To verify the effectiveness of our attack, we then mount 100 key-recovery attacks under 100 random master keys and get the $100\%$ attacking success rate. 
The remaining 32-bit information of mask key can be recovered by brute-force attacks. Consequently, the time complexity of this 14-round key-recovery attack is $2^{32}$ and the data complexity is $2^{10}\times 28\times 2 = 2^{15.81}$ chosen plaintexts.

\vspace{4pt}
\item For 15-round key-recovery attack, we use the 14-round RX-neural distinguisher as shown in Table~\ref{table:staged_result}. It is trained using half RX-difference $[4, 0x22]$ and data format {$\mathscr{D}_{5}^{28}$}, with a accuracy of 0.5141. The corresponding 13-round one has the acuracy of 0.7105. The parameters of this attack are set as
\begin{center}
\renewcommand\tabcolsep{8pt}
\renewcommand{\arraystretch}{1.5} 
\begin{tabular}{llll}
                    $m = 2^{10}$ & \hspace{0.1cm} $k=28$ & \hspace{0.1cm} $c_{1} = 16$ & \hspace{0.5cm} $c_{2} = 500$ \\
                    $t = 2^{10}$ & \hspace{0.1cm} $l = 6$ & \hspace{0.1cm} $n = 32$. & 
                \end{tabular}
            \end{center}
Fig.~\ref{fig:simon32_mean_data25_15r_pairs28} shows the WKR profile of the 15-round subkey. The analysis reveals that the correct key exhibits a mean response value of 0.504, closely aligning with the accuracy of 14-round RX-neural distinguisher. However, this differentiation is not as statistically significant as observed in the 14-round WKS, therefore theoretically yielding lower success rates compared to the 14-round attack. Empirical verification via 100 key-recovery attacks demonstrate an achievable success rate of $75\%$. The time and data complexities are the same as the 14-round attack. 
\end{itemize}

\section{Key-recovery Attacks on Simeck32/64}
\label{sec:6_0}
In related-key setting, Simon's RX-difference of subkey pair for each round is uniquely determined as its key schedule is linear. If one subkey of the pair is guessed, then another one can be directly obtained using the deterministic RX-difference. However, Simeck's RX-difference is probabilistic due to its nonlinear key schedule, requiring simultaneous guessing both of the two subkeys when mounting key-recovery attacks under related-key setting. Therefore, BKS is not suitable for Simeck as BKS only supports single-key guessing. 

\vspace{5pt}
\noindent\textbf{The joint wrong key response.} In~\cite{bao2023more}, Bao et al. used fixed key differences in related-key differential attacks against Speck, known as weak-key attacks. It makes sense but can only success under a fixed key space (weak-key space) instead of the full key space. To ensure that key-recovery attacks can be achievable for the full key space under related-key setting, we combine the idea of joint distribution to construct the wrong key response for key pairs, called the \textit{joint wrong key response} (JWKR). 
When attacking Simon32/64, the WKR profile is constructed by systematically enumerating all possible values of the target subkey (comprising $2^{16}$ candidates) and computing their corresponding mean and standard deviation responses. This process requires approximately 25 minutes. Extending this methodology to JWKR profile construction, which involves applying the same exhaustive approach to the full $2^{32}$ subkey pair space, yields a mathematically projected runtime of $25\times 2^{16}\approx 2^{20.6}$ minutes (equivalent to three years). Obviously, it is unfeasible. To tackle this problem, we introduce \textit{key bits sensitivity test} (KBST) method aiming to reduce the subkey pair space of JWKR.

\begin{figure*}[!h]
            \centering
            \subfloat[For 15-round subkey]{\includegraphics[width=2.5in]        {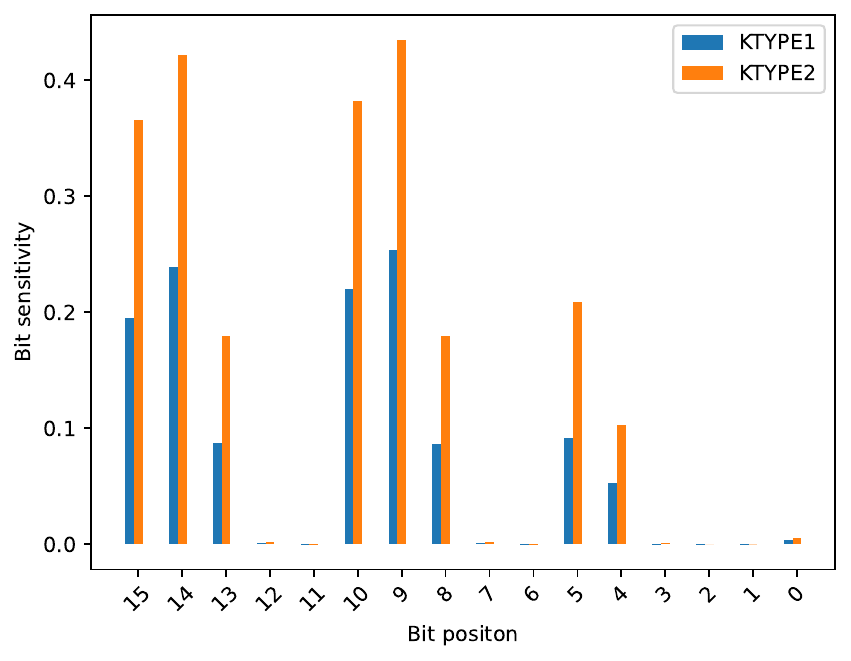}}
            \label{fig:simeck_r16pairs36_keybst}
            \hspace{2cm}
            \subfloat[For 16-round subkey]{\includegraphics[width=2.5in]{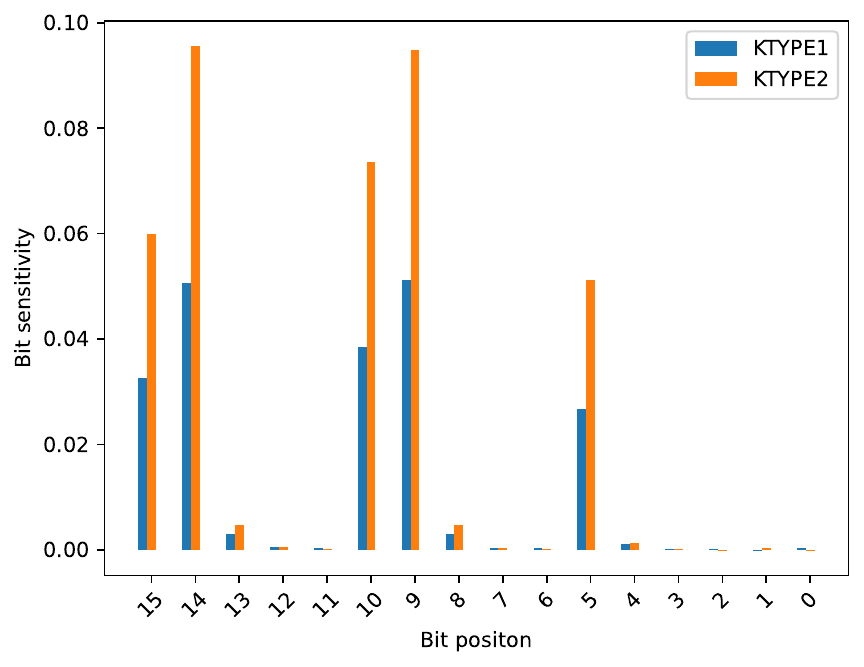}}
            \label{fig:simeck_r17pairs36_keybst}
            \caption{The key bits sensitivity test for Simeck32/64}
            \label{fig:6simeck_keybst}
	\end{figure*}

\vspace{5pt}
\noindent\textbf{Key bits sensitivity test.} Inspired by the application of BST to detecting the redundant bits of ciphertext (as illustrated in Sect.~\ref{sec:4_1}), we can use BST to detect the influence of each bit of subkey on the accuracy of trained RX-neural distinguisher. If the key bits have a negligible influence on the accuracy, then we call these bits as insensitive key bits, otherwise as sensitive key bits. To distinguish from BST for ciphertext, we call this \textit{key bits sensitivity test} (KBST). We take the bit position $0$ of $16$-round subkey of Simeck32/64 as an instance to illustrate the process of KBST. To launch KBST for $16$-round subkey, we need to prepare a $15$-round RX-neural distinguisher. We choose the one with accuracy 0.6042 trained using half RX-difference $[1, 0x4]$ and data format $\mathscr{D}_{7}^{36}$. Firstly, we randomly choose $10^6\times 36$ pairs of plaintext satisfing the half RX-difference $[1, 0x4]$ and generate a set composed of $10^{6}\times 36$ pairs of ciphertext, denoted by $\mathbb{C}$. Then we choose $10^{6}$ masks where the value corresponding to the bit position 0 varies randomly, and the values of other bit positions are all 0s. We XOR these masks one by one on the fixed 16-round subkey pair simultaneously to generate $10^6$ modified subkey pairs. Decrypt all elements in $\mathbb{C}$ using the modified subkey pairs and combine with data format $\mathscr{D}_{7}^{36}$, we obtain a validation dataset composed of $10^6$ samples\footnote{Note that $\mathbb{C}$ contains $10^6 \times 36$ pairs of ciphertext but there are only $10^6$ modified pairs of subkey, we have to divide $\mathbb{C}$ into $10^6$ groups. Each group containing 36 pairs of ciphertext is decrypted using one modified subkey pair, and the corresponding decrypted group forms one sample using the 36-multi-ciphertext data format $\mathscr{D}_{7}^{36}$.}.
Send this dataset to the prepared 15-round neural distinguisher, a validation accuracy is returned. The difference of this validation accuracy and the original accuracy of $15$-round neural distinguisher is regarded as the $\mathtt{KTYPE1}$ sensitivity of the $0$-th bit of 16-round subkey. As for the another type of bit sensitivity, denoted by $\mathtt{KTYPE2}$, is derived using the constant mask where the value corresponding to $0$-th bit is 1 and others are 0s. Considering the other 15 bit positions, KBST of the 16-round subkey can be accomplished and the result is exhibited in Fig.~\ref{fig:6simeck_keybst}. It appears that only five sensitive key bits in 16-round subkey, namely, it is sufficient to construct JWKR for these five bits. The KBST for more rounds and data formats can be found in Sect.~IV of Supplemental Material. It is worth noting that KBST of 18-round subkey is similar to that of other rounds subkeys at the positions $[4, 5, 8, 9, 10, 13, 14, 15]$ bits, confirming that our 17-round RX-neural distinguisher for Simeck32/64 is valid, even though it is very weak.

 \begin{figure*}[!h]
		\centering
		\subfloat[For 15-round subkey]{\includegraphics[width=2.5in]{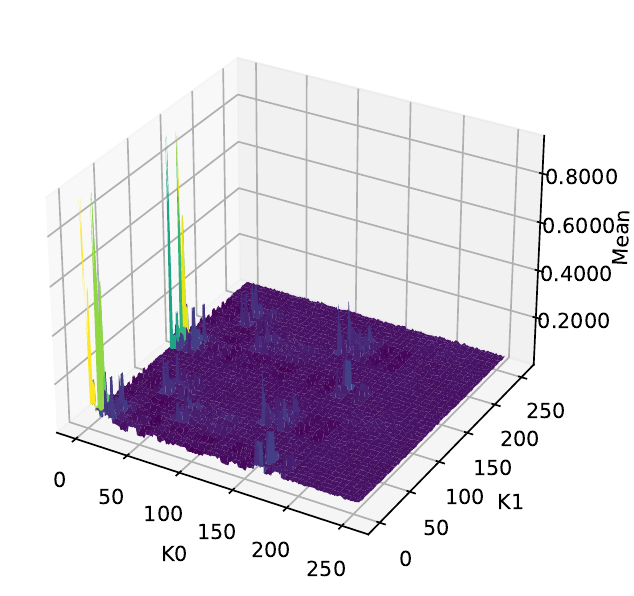}}
		\label{fig:6simeck_r152_wrong}
            \hspace{1cm}
		\subfloat[For 16-round subkey]{\includegraphics[width=2.5in]{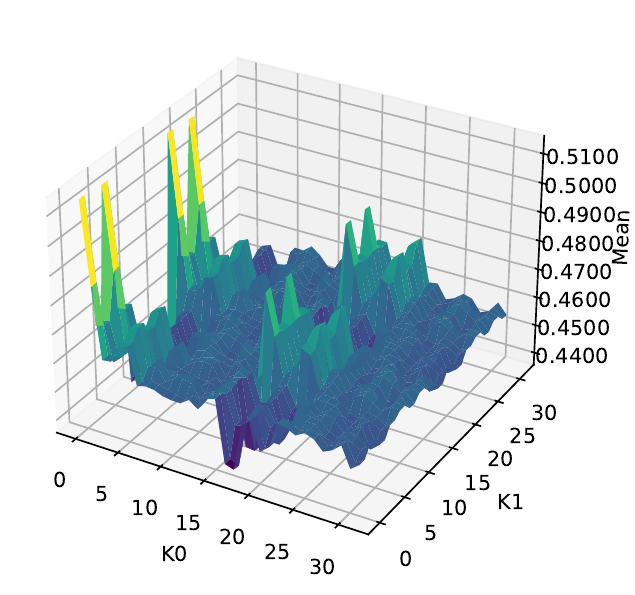}}
		\label{fig:6simeck_r162_wrong}
		\caption{The joint wrong key response for sensitive bits of Simeck32/64
              }
		\label{fig:6simeck_wrong}
	\end{figure*}
 \begin{figure*}[!h]
		\centering
    	\subfloat[KBST]{\includegraphics[width=2.5in]{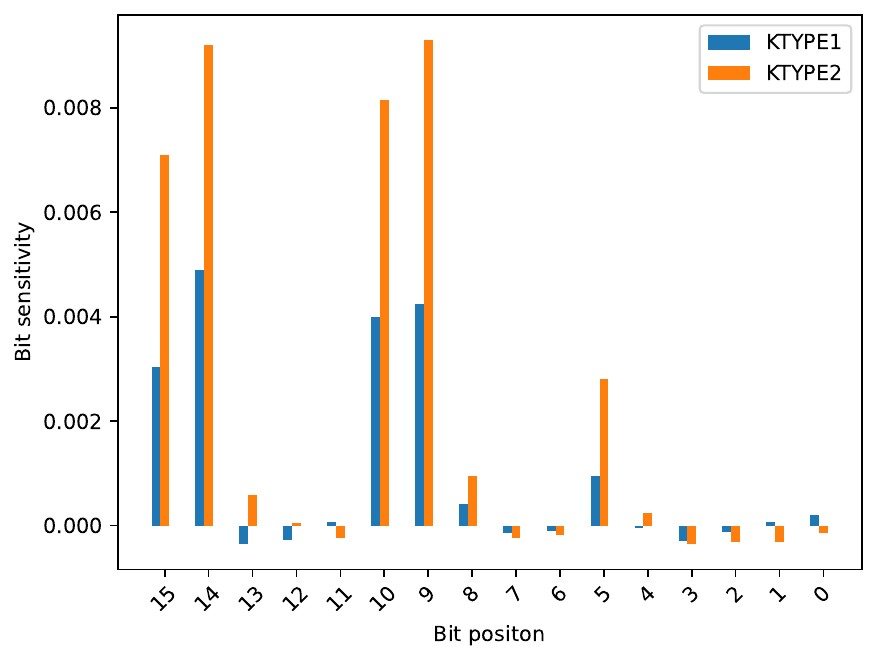}
         \label{fig:6simeck_r1722_bst}}
            \hspace{1cm}
		\subfloat[JWKS]{\includegraphics[width=2.5in]{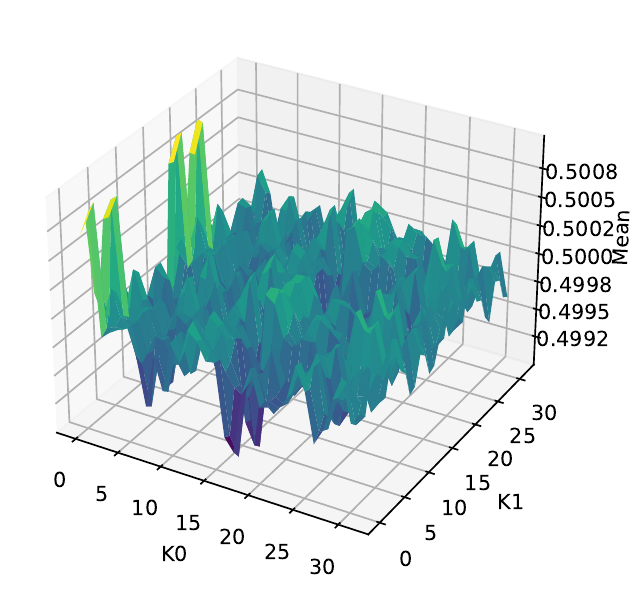}\label{fig:6simeck_r1722_wrong}}
		
		\caption{The KBST and JWKS for 17-round Simeck32/64}
		\label{fig:6simeck_wrong_kbst}
	\end{figure*}

\noindent\textbf{Key-recovery attacks on 16- and 17-round Simeck32/64.} Key-recovery attacks for Simeck can also make use of the attack framework as described in Sect.~\ref{sec:5_0}, but need to perform JWKR instead of WKR in BayesianKeySearch algorithm. Therefore, we here reuse some notations denoted in Simon's key-recovery attacks. All the RX-neural distinguishers utilized in 16- and 17-round key-recovery attacks are trained using half RX-difference [1, 0x4] and data format $\mathscr{D}_{7}^{36}$. The details of attacks are illustrated as follows:
\begin{itemize}
    \item For 16-round attack, we utilize the 14- and 15-round distinguishers with respective accuracies of 0.8982 and 0.6042. The KBSTs for 15- and 16-round subkeys are shown in Fig.~\ref{fig:6simeck_keybst}. It indicates that we need to construct JWKRs for bit positions $[4,5,8,9,10,13,14,15]$ of $rk^{15}$ and $[5,9,10,14,15]$ of $rk^{16}$, which are illustrated by three-dimensional images in Fig~\ref{fig:6simeck_wrong}. The JWKS for more rounds can be found in Sect.~V of Supplemental Material. By setting the attack parameter as
        \begin{center}
            \renewcommand\tabcolsep{8pt}
            \renewcommand{\arraystretch}{1.5} 
                \begin{tabular}{llll}
                    $m = 2^{10}$ & \hspace{0.1cm} $k=36$ & \hspace{0.1cm} $c_{1} = 150$ & \hspace{0.5cm} $c_{2} = 1000$ \\
                    $t = 2^{10}$ & \hspace{0.1cm} $l = 4$ & \hspace{0.1cm} $n = 32$, & 
                \end{tabular}
            \end{center}

we can straightforwardly recover the 13 bits of $(rk^{15}, rk^{16})$ within seven minutes and $2^{16.17}$ chosen plaintexts, and the success rate is $98\%$. The remaining 51-bit information can be recovered by brute-force attack. Thus the time complexity is $2^{51}$.

 \vspace{4pt}
 \item For 17-round attack, we use the 15- and 16-round neural distinguisher with the accuracies of 0.6042 and 0.5130, repsectively. According to KBST of $rk^{17}$ as shown in Fig.~\ref{fig:6simeck_r1722_bst}, there are also five sensitive bits: $[5,9,10,14,15]$, and the corresponding JWKS is exhibited in Fig~\ref{fig:6simeck_r1722_wrong}. The attack parameters are set as:
  \begin{center}
    \renewcommand\tabcolsep{8pt}
    \renewcommand{\arraystretch}{1.5} 
                \begin{tabular}{llll}
                    $m = 2^{10}$ & \hspace{0.2cm} $k=36$ & \hspace{0.2cm} $c_{1} = 4$ & \hspace{0.5cm} $c_{2} = 100$ \\
                    $t = 2^{10}$ & \hspace{0.2cm} $l = 4$ & \hspace{0.2cm} $n = 32$. & 
                \end{tabular}
            \end{center}
As a consequence, 10 bits of $(rk^{16},rk^{17})$ can be recovered with data complexity $2^{16.17}$. The whole process takes about four hours, and the success rate is about $40\%$. The total time complexity to recover 64-bit key is $2^{54}$.
\end{itemize}

\section{Conclusion}
\label{sec:7_0}
In this paper, we conducted a comprehensive study on the deep learning-based rotational-XOR cryptanalysis for Simon32/64 and Simeck32/64. Specifically, inspired by previous work about the differential-neural attack, we proposed two fundamental data formats specially designed for training RX-neural distinguishers. Through systematic exploration of initial RX-differences with Hamming weight $\leq 2$, we identified some good input patterns that enable the development of high-accuracy RX-neural distinguishers. In addition, we further enhanced the performance of the trained neural distinguishers using multi-ciphertext data formats and staged training techniques. Finally, to verify effectiveness of the proposed RX-neural distinguishers, some round-reduced key-recovery attacks were presented. 

Regarding the neural-based distinguishers, our results extend the state-of-the-art from 13 to 14 rounds for Simon32/64 and from 15 to 17 rounds for Simeck32/64. 
Our key-recovery attacks under the related-key setting do not surpass existing single-key differential-neural attacks~\cite{zhang2022improving} and~\cite{zhang2023improved} as we cannot prepend additional rounds in front of related-key neural distinguisher, but this work establishes two notable contributions: 1) present the first key-recovery attacks for both cipher in related-key setting. 2) develop two novel techniques, named \textit{key bits sensitivity test} and \textit{joint wrong key response}, effectively addressing challenges in applying neural distinguishers to related-key attacks without considering weak-key space. These advancements provide new insights into RX-neural cryptanalysis, particularly in related-key attacks based on neural distinguishers, and it is also potential to explore better attacks by combining with other types of related-key neural attacks like related-key differential-neural attack.

\section*{Acknowledgement}
This work was supported by the National Natural Science Foundation of China (No. 62272147, 12471492, 62072161, 12401687) and the Innovation Group Project of the Natural Science Foundation of Hubei Province of China (No. 2023AFA021).

\bibliographystyle{IEEEtran}
\bibliography{ref_reduced.bib}

\begin{IEEEbiographynophoto}{Chengcai Liu}
			received the M.E. degree with School of Cyber Science and Technology in Hubei University, Wuhan, China. His research interest includes cryptanalysis of block ciphers.
		\end{IEEEbiographynophoto}
  \vspace{-25pt}
\begin{IEEEbiographynophoto}{Siwei Chen} received the Ph.D. degree from Hubei University in 2022. He is currently a Lecturer with School of Cyber Science and Technology in Hubei University, Wuhan, China. His current research interests include design and cryptanalysis of symmetric ciphers.
		\end{IEEEbiographynophoto}
  \vspace{-25pt}
		\begin{IEEEbiographynophoto}{Zejun Xiang} received the Ph.D. degree from University of Chinese Academy of Sciences, Beijing, China, in 2018. He is currently an Associate Professor with School of Cyber Science and Technology in Hubei University, Wuhan, China. His current research interests include design, cryptanalysis, classical and quantum implementation of symmetric ciphers.
		\end{IEEEbiographynophoto}
 \vspace{-25pt}
        \begin{IEEEbiographynophoto}{Shasha Zhang} received the Ph.D. degree from Wuhan University, Wuhan, China, in 2009. She is currently an Associate Professor with School of Cyber Science and Technology in Hubei University, Wuhan, China. Her current research interests include cryptanalysis, classical and quantum implementation of symmetric ciphers and blockchain.
		\end{IEEEbiographynophoto}
  \vspace{-25pt}
		\begin{IEEEbiographynophoto}{Xiangyong Zeng} received the Ph.D. degree from Beijing Normal University, Beijing, China, in 2002, then did the post-doctoral work from 2002 to 2004 in Wuhan University, Wuhan, China. He is currently a Professor with Faculty of Mathematics and Statistics in Hubei University, Wuhan, China. His current research interests include coding theory, cryptology, and blockchain.
		\end{IEEEbiographynophoto}
\vfill
\end{document}